\documentclass[a4paper,11pt]{article}
\pdfoutput=1 

\usepackage{jcappub} 

\usepackage[T1]{fontenc} 

\usepackage{graphicx}
\usepackage{amssymb}
\usepackage{ulem}
\usepackage{multirow}
\usepackage{color}
\usepackage{rotating}
\usepackage{placeins} 
\usepackage[ampersand]{easylist}

\usepackage{subfig}

\def\be{\begin{equation}}
\def\ee{\end{equation}}

\def\bi{\begin{itemize}}
\def\ei{\end{itemize}}

\def\ben{\begin{enumerate}}
\def\een{\end{enumerate}}

\def\bt{\begin{tabular}}
\def\et{\end{tabular}}

\def\bc{\begin{center}}
\def\ec{\end{center}}

\newcommand{\bes}{\begin{subequations}}
\newcommand{\ees}{\end{subequations}}
\def\bea{\begin{eqnarray}}
\def\eea{\end{eqnarray}}

\def\tcr{\textcolor{red}}

\newcount\hh
\newcount\mm
\mm=\time
\hh=\time
\divide\hh by 60
\divide\mm by 60
\multiply\mm by 60
\mm=-\mm
\advance\mm by \time
\def\hhmm{\number\hh:\ifnum\mm<10{}0\fi\number\mm}

\title{Primordial non-Gaussianities in  single field inflationary models with non-trivial initial states}

\author{Sina Bahrami and \'Eanna \'E. Flanagan}
\affiliation{Department of Physics, Cornell University, Ithaca, NY 14853, USA.}

\affiliation{\tcr{Draft of August 19, 2014; printed \today{} at \hhmm}}

\emailAdd{sb933@cornell.edu}
\emailAdd{eef3@cornell.edu}

\abstract{
We compute the non-Gaussianities that arise in single field, slow roll
inflationary models arising
from arbitrary homogeneous initial states, as well as subleading
contributions to the power spectrum.
Non Bunch-Davies vacuum initial states can arise if the transition to
the single field, slow roll inflation phase occurs only shortly before
observable modes left the horizon.
They can also arise from new physics at high energies that has been
integrated out.
Our general result
for the bispectrum exhibits several features that were previously
seen in special cases.

}

\begin{document}
\maketitle
\flushbottom

\section{Introduction and summary}\label{intro}

\subsection{Background and motivation}\label{context}

Inflationary models are in good quantitative agreement with data
obtained from the cosmic microwave background (CMB) and large scale
structure \cite{weinberg,dodelson,baumann,mukhanov,
planck14,planck15,planck21,planck23}.
Present experimental data already rule out some inflationary models
and constrain others. The CMB temperature inhomogeneities
are distributed following a nearly scale-invariant spectrum, which is
consistent with the predictions of inflation
\cite{mukhanov1981quantum,1982PhLB..115..295H,1982PhRvL..49.1110G,starobinsky117dynamics,1983PhRvD..28..679B}.
In addition, one of the key predictions of
inflation is that the statistical properties of the fluctuations are
Gaussian to high accuracy, which has been confirmed by observations,
most recently by the Planck satellite \cite{planck23}.
However, small amounts of non-Gaussianity are predicted to be present, with a size that depends on the details of the model, and  it is hoped to use future observations of non-Gaussianity to differentiate between different models \cite{komael}.

The robustness of the predictions of inflation depends in part on the
assumption that inflation started considerably before the horizon exit
of the largest modes we can observe today. This occurs in any model
where the total number $N$ of $e$-folds of inflation is significantly
larger than the minimum number $N_{\text{min}} \approx 60$ required to
solve the horizon and flatness problems. In many models $N \gg
N_{\text{min}}$ is very natural, while $N \approx N_{\text{min}} +
\text{ (a few)} $ is somewhat unnatural or fine tuned. Nevertheless it
is still interesting to consider the possibility that $N \approx
N_{\text{min}} + \text{ (a few)} $, since it opens up a richer set of
possibilities for inflationary predictions. In particular, the well
known statistical anomalies of low multipoles of the CMB, as recently
confirmed and extended by Planck \cite{planck22}, may be a hint in
this direction.

When the epoch of inflation that we can probe observationally occurs shortly after the start of inflation, there are a number of different non-standard effects that can arise. One is non-isotropy of the geometry: the initial geometry can have a Kasner-like anisotropic component that leaves an imprint on the perturbations at late times. The leading order effect on the power spectrum in these models can be written as
\be
P(\vec{k}) = P_0 (k) + P_1 (k) Q_{i j } \hat{k} ^{i} \hat{k}^{j} ,
\label{1.1}
\ee
where ${\vec k}$ is a spatial wavevector, $ k = | \vec{k}| $ is its
magnitude and $\hat{k}^{i} = k^{i}/k$.  Here $Q_{i j}$ is a symmetric
traceless tensor which singles out preferred directions in space
associated with the anisotropy at early times, and the functions $P_0$
and $P_1$ are functions of $k$ only. These models have
been studied by
Refs.\ \cite{pitrou,kim,gumr,samal,aluri,pullen,dey0,dey}. However, the
recent Planck data show no
statistically significant quadrupolar component of the form
(\ref{1.1}) [see Fig.\ 34 of Ref.\ \cite{planck22}].

A second type of effect that can occur when inflation starts just
before the epoch we observe is non Bunch-Davies vacuum initial states.
These can arise from effects at earlier times that are
outside the domain of validity of a single field model, for example
multifield effects (see eg. Ref.\ \cite{mcallister}).
The choice of initial state is constrained somewhat by the requirement
that backreaction due to the energy density and pressure of the
initial state must be a small perturbation to the inflationary
background, in order to have a self consistent computational
framework, see section \ref{sec:back} below.

Non Bunch-Davies initial states can modify both the spectrum and
bispectrum of the perturbations, and in particular can enhance the
size of the bispectrum. Non-Gaussianities arising from specific
classes of nontrivial initial states have been studied by many
authors, including particle number eigenstates \cite{Parker},
Gaussian states \cite{agarwal},
general multimode squeezed (i.e. vacuum) states
\cite{porrati1,porrati2,holt,meer,meer1,ganc,shandera,Ashoorioon:2013eia,Aravind:2013lra}, coherent states
\cite{kundu}, thermal states \cite{das}, and so-called "calm states"
\cite{ashoorion}.
In this paper we shall generalize these results to arbitrary initial
states that are homogeneous.
  As first
pointed out by Agullo and Parker \cite{Parker}, the squeezed triangle limit of the bispectrum can be
significantly enhanced over its Bunch-Davies value; this remains true
for the more general initial states analyzed here.

A second motivation for considering non Bunch-Davies vacuum
initial states, separate from the possibility of inflation starting
shortly before the epoch probed by current observations,
is provided by the trans-Planckian issue
\cite{goldstein,brandenmartin,collinsholman,egks,egks1,brandenmartin01,bh,lmmp,egks02,daniel,kk}.
Modes
that we observe today have physical momenta $k_{\text{phys}}=
k/a$ which exceeds the Planck scale $M_{p}$ at sufficiently early times.
A complete description of the
physics at that stage will only be possible once we have a solid
understanding of quantum gravity.  For a given mode, the quantum
gravitational corrections to the dynamics at horizon crossing are suppressed
by $\sim H^2/M_p^2$ if the evolution is always adiabatic, where $H$ is the
Hubble parameter during inflation, but could be
larger and scale as $\sim H / M_p$ if violations of adiabaticity occur.
In many toy models of quantum gravity effects, the effects of the
Planck scale corrections to the dynamics of the modes can be
mimicked by using the standard dynamics but with a modified initial
state.  Many computations have been performed of
modifications to the spectrum and bispectrum of the perturbations
due to new physics at high energies \cite{kk,egks,egks1,brandenmartin01,bh,lmmp,egks02,daniel,Ashoorioon:2013eia}.
The general results of this paper could be used as a tool in such
computations.

\subsection{Summary of results}\label{results}

In this paper we start with a general mixed quantum state
describing a statistical distribution of initial inflaton
perturbations, constrained only by the assumption of spatial
homogeneity. We calculate the scalar power spectrum and bispectrum of the
comoving curvature perturbation ${\cal R}$.
Our general result
for the bispectrum, given in Eq.\ (\ref{finalresult}) below,
exhibits several features that were previously
seen in special cases:
\begin{itemize}

\item The three point function of the initial
state will generically\footnote{By ``generic'' we mean that there is
  no suppression of the initial three point function compared to the
  initial two and four point functions.  This can be violated in
  specific scenarios, as discussed further in Sec.\ \ref{sec:genericstates} below.}
give the dominant contribution to the
bispectrum, since the contributions from the initial two and four
point functions are suppressed by the factor $\sim H \sqrt{\epsilon}/M_p$,
where $H$ is the inflationary Hubble scale,
$M_p$ is the Planck scale, and $\epsilon$ is a slow roll parameter.
This dominance of the three point function was previously seen in a
special class of states by Agarwal {\it et al.} \cite{agarwal}.
It is easy to understand: a nonzero bispectrum is obtained from an
initial three point function from just the linear evolution,
whereas contributions from the
initial two and four point functions require nonlinearities.
The bispectrum can be significantly larger than the Bunch-Davies
bispectrum, cf.\ Eq.\ (\ref{enhanced00}) below,
even when initial
occupation numbers are small and corrections to the power spectrum
[Eq.\ (\ref{ps0}) below] are
small.

\item For initial states with vanishing three point function (the
subject of most previous studies),
the bispectrum can be enhanced compared to the Bunch-Davies bispectrum
in the squeezed triangle limit, as discovered by Agullo and Parker
\cite{Parker} for initial states consisting of statistical mixtures of
occupation number eigenstates.  For initial occupation numbers of
order unity, the enhancement factor is the ratio
of wavenumbers of the large scale mode and the small scale mode [Eq.\
(\ref{large}) below].  We
argue that backreaction considerations limit the enhancement factor to
be $\lesssim 200$ (Sec.\ \ref{sec:lsr}).

\item For initial states with vanishing three
point function, the bispectrum can be enhanced compared to the
Bunch-Davies bispectrum in the elongated triangle limit,
as discovered by Chen  {\it et al.} \cite{Chen:2006nt} and Holman and Tolley \cite{holt} for generalized vacuum
states (multimode squeezed states), and studied extensively by Agullo
and Shandera \cite{shandera}.
We argue that the enhancement factor is limited by backreaction
considerations to be of the same order as the enhancement factor of
the squeezed triangle limit or smaller (Sec.\ \ref{sec:lsr}).

\item The
bispectrum for non-vacuum initial states can violate the consistency
relation of Creminelli and Zaldarriaga \cite{cre-zald} for the squeezed triangle
limit, which assumes an initial Bunch-Davies vacuum, as noted
previously in Refs.\ \cite{agarwal,shandera}.

\end{itemize}

\section{Review of basic results in single field inflationary models}

\subsection{Generation of perturbations neglecting interactions}

The action for single field slow-roll models of inflation is
\be
 S = \frac{1 }{2} \int d^4 x   \sqrt{-g} [ M_p^2 R- (\bigtriangledown \phi)^2 - 2 V (\phi) ],
\label{2.1}
\ee
where $R$ is the Ricci curvature, $\phi$ is the inflaton field,
and we use units with $c = \hbar=  1 $.
We consider linear scalar perturbations about a background
Friedmann-Robertson-Walker metric in comoving gauge, where
\bes
\label{2.2}
\bea
ds^2 &=& -(1 +2 \Phi) dt^2 +a(t)^2 (1-2 \Psi) dx^2 + 2 a(t)  B_{,i}
dx^{i} dt , \\
\phi(t, \vec{x})&=& \phi (t) + \delta \phi(t, \vec{x}),
\eea
\ees
with $ \delta \phi(t, \vec{x}) =0 $ \cite{baumann}. Here $a(t)$ is the scale factor and $\Phi$, $\Psi$, and $B$ are scalar potentials. The background equations of motion are the Friedmann equations
\be
3 M_p^2 H^2 = \frac{\dot{\phi}^2}{2} + V(\phi) , \hspace{2cm} \dot{H}=- \frac{1}{2 M_p^2} \dot{\phi}^2   , \ee
where $H \equiv \dot{a}/{a}$ is the Hubble parameter, and the Klein Gordon equation
\be \ddot{\phi}+ 3 H \dot{\phi} +  V'(\phi)=0 .
\ee
The potential slow roll parameters $\epsilon$ and $\eta$ are defined by
\be
\epsilon = \frac{M_p^2}{2} \left(\frac{V'}{V} \right)^2 \approx \frac{{\dot \phi}^2}{2 H^2 M_p^2}, \hspace{2cm}
\eta = M_p ^2 \left( \frac{V''}{V} \right) \approx - \frac{{\ddot \phi}}{H {\dot \phi}}
+ \frac{{\dot \phi}^2}{2 H^2 M_p^2},
\ee
where prime denotes differentiation with respect to the field $\phi$, and the approximate equalities are valid in the field slow roll limit.

We define a gauge invariant quantity called the comoving curvature
perturbation $\mathcal{R}$ by
\be
\mathcal{R} = \Psi + \frac{H}{\dot{\phi}} \delta \phi.
\ee
In our choice of gauge we have $\mathcal{R} = \Psi $.
To find the leading order action describing the evolution of $
\mathcal{R}$ we expand the  action (\ref{2.1}) to second order in
$\mathcal{R}=\Psi$, $B$ and $\Phi$, integrate out the non-dynamical
fields $\Phi$ and $B$, and simplify using the background equations of
motion and using integrations by parts.  The final result is
\cite{baumann}
\be
S_2= \frac{1}{2} \int d^3x \int d \tau z^2 [ \mathcal{R}'^2 - (\partial{\mathcal{R}})^2  ], \
\label{2.6}
\ee
where $ \tau \equiv \int dt/a(t) $ is the conformal time, $z \equiv a
\dot{\phi} / H$, primes denote derivatives with respect to $\tau$, and
$ (\partial \mathcal{R})^2 \equiv \delta^{ij} \mathcal{R}_{,i}
\mathcal{R}_{,j}$.
We decompose the curvature perturbation into spatial Fourier modes via
\be
 \mathcal{R}(\vec{x},\tau) = \int \frac{d^3 k}{(2 \pi )^3 }
 \mathcal{R}_{\vec{k}}(\tau) e ^{i \vec{k} . \vec{x}}.
\label{2.7}
\ee
Using the decomposition (\ref{2.7}) in the equation of motion obtained
from the action (\ref{2.6}) yields
the Mukhanov-Sasaki equation 
\be
\mathcal{R}_{\vec k}''(\tau) + \frac{2}{z} z' \mathcal{R}_{\vec k}'(\tau) + k^2
\mathcal{R}_{\vec k}(\tau)=0 .
\label{2.8}
\ee
In the slow-roll limit, we have $ z^{-1} z' \approx (1+ 3 \epsilon -
\eta ) / \tau$.

We now quantize the theory by promoting $\mathcal{R}_{\vec k}(\tau)$
to an operator,
$\mathcal{R}_{\vec k}(\tau)  \rightarrow
\hat{\mathcal{R}}_{\vec k}(\tau)$.
Next for each $k$ we choose any complex solution $\mathcal{R}(k,\tau)$
of the Mukhanov-Sasaki equation (\ref{2.8}), for which ${\cal R}(k,\tau)$ and
${\cal R}(k,\tau)^*$ form a basis of the two dimensional solution space
and for which the quantity (\ref{2.10}) below is positive.
Since the operator ${\hat {\cal R}}_{\vec k}$ satisfies the Mukhanov-Sasaki
equation we can decompose it on this basis:
\be
{\hat {\cal R}}_{\vec k}(\tau) =  \hat{A}_{\vec{k}}
\mathcal{R}(k,\tau) +\hat{ A} ^{\dagger}_{-\vec{k}} \mathcal{R}
(k,\tau)^* ,
\label{2.9}
\ee
where $ \hat{A}_{\vec{k}} $ and $ \hat{A}^{\dagger} _{\vec{k}}$ are
annihilation and creation operators.  We normalize the mode functions
by requiring
\be
i z^2 (\mathcal{R}^{*} \mathcal{R}' - \mathcal{R}'^*
 \mathcal{R}) =1 .
\label{2.10}
\ee
It then follows from Eq.\ (\ref{2.6}) that $ [ \hat{A}_{\vec{k}} ,
\hat{A}^{\dagger} _{\vec{k}'}]= (2 \pi)^3 \delta (\vec{k}-\vec{k}')
$.

This construction allows some freedom in the choice of mode function
${\cal R}(k,\tau)$, which can be resolved by specifying a boundary
condition at early times.  The choice of mode function determines a
corresponding choice of a vacuum state $|0\rangle $ for which $
\hat{A}_{\vec{k}} | 0 \rangle = 0 $. The standard choice of vacuum is
known as the Bunch-Davies vacuum \cite{bunchdavies}, which is the
Minkowski vacuum of a comoving observer in the distant past
when the mode is deep inside the horizon. This choice
of vacuum corresponds to the following boundary condition on the mode
function
\cite{baumann}
\be \mathcal{R}(k,\tau) \rightarrow \frac{e^{-i k \tau}}{z(\tau) \sqrt{2 k}} , \hspace{1 cm} \text{for } |k \tau | \gg 1 .  \ee
With this in hand and assuming the slow-roll parameters to be constant, we can find the unique solution to equation (\ref{2.8}) \cite{baumann}
\be
\mathcal{R}(k,\tau) =\Big ( \frac{- \pi \tau}{4  z^2 } \Big )^{1/2} H ^{(1)} _{\mu} (-\tau k),
\label{3.13}
\ee
where $H ^{(1)} _{\mu} $ is a Hankel function with index $\mu = 3/2 +
2 \epsilon - \eta $. In the superhorizon regime $ |\tau k | \ll 1 $,
and assuming $\epsilon \ll 1$ and $\eta \ll  1$, the mode function
becomes
\be
\mathcal{R}(k,\tau) \rightarrow  i\left[ \frac{1}{4  k^3
  \epsilon(\tau_k) }\right]^{1/2} \frac{H(\tau_k)}{M_p}.
\label{2.13}
\ee
Here $\tau_{k}$ is the value of conformal time at which the mode $k$
exits the horizon, given by $ k= a(\tau_k) H$.

\subsection{Including the leading order effects of interactions}
\label{sec:interactions}

The leading order nonlinearities for inflaton perturbations were first
worked out in detail by Maldacena \cite{maldacena}.
The interaction Hamiltonian is
obtained by expanding the action (\ref{2.1}) to third  order in $
\mathcal{R}$ and integrating out the non-dynamical
fields (the lapse and the shift). Maldacena obtained a
simple form for the interaction Hamiltonian by writing it in terms of
a redefined curvature field $\mathcal{R}_c$ which is given by
\be
\mathcal{R} = \mathcal{R}_c + \Big[ \frac{1}{2}
\frac{\ddot{\phi}}{\dot{\phi} H} + \frac{1}{8 M_p ^2} \frac{\dot{\phi}
  ^2} {H^2} \Big ] \mathcal{R}_c ^2  + \frac{1}{4 M_p ^2}
\frac{\dot{\phi}^2} {H^2} \partial^{-2} (\mathcal{R}_c \partial^2
\mathcal{R}_c ).
\label{5.2}
\ee
Expanding the action
(\ref{2.1}) around the spatially homogeneous background solution to
third order in a specific choice of gauge, Maldacena found
\be
S[\mathcal{R}] = S_2[\mathcal{R}_c] + S_3[\mathcal{R}_c] + \ldots,
\ee
where the functional $S_2$ is the quadratic action (\ref{2.6})
evaluated at ${\cal R} = {\cal R}_c$, and
$S_3$ is given by
\be
S_3(\mathcal{R}_{c} ) = - \int d \tau H_{\rm int} = \int d \tau \int d^3x \, a(\tau)^3
\Big( \frac{\dot{\phi}}{H} \Big)^4 H M_{p} ^{-2} \mathcal{R}_{c}
'^2 \partial^{-2} \mathcal{R}_{c}'.
\label{inter}
\ee
Here as before primes denote differentiation with respect to $\tau$,
and $H_{\rm int}$ is the interaction Hamiltonian.  We will use this
interaction Hamiltonian and the field redefinition (\ref{5.2})
to compute corrections to the spectrum in section
\ref{sec:power} below, and to compute the bispectrum in section
\ref{sec:bispectrum} below, for general initial states.

In the remainder of this paper, we shall perform computations using
the redefined field ${\cal R}_c$, and use the field redefinition only
to compute the observable quantities that are defined in terms of
${\cal R}$.  In
particular, the mode creation and annihilation operators will
from now on be defined in terms of a mode expansion of the field
${\hat {\cal R}}_c$:
\be
{\hat {\cal R}}_{c,{\vec k}}(\tau) =  \hat{A}_{\vec{k}}
\mathcal{R}(k,\tau) +\hat{ A} ^{\dagger}_{-\vec{k}} \mathcal{R}
(k,\tau)^*.
\label{2.9a}
\ee

\section{Choice of initial state}
\label{sec:initial}
\subsection{Parameterization of general homogeneous states}
\label{sec:notation}

In this paper we will allow arbitrary homogeneous initial
states,
a class of states more general than that considered by Agullo and
Parker \cite{Parker}, who assumed statistical mixtures of particle
number eigenstates.  We now turn to a description of how the initial
state is parameterized.

For an initial density matrix $\hat{\rho}$, the spectrum
and bispectrum of ${\cal R}$ at late times will be determined by the
initial two point, three point, and four point functions at some initial
conformal time $\tau_0$, $\langle
\hat{A}^{\alpha} _{\vec{k}_1} \hat{A}^{\beta} _{\vec{k}_2} \rangle$,
$\langle \hat{A}^{\alpha} _{\vec{k}_1} \hat{A}^{\beta} _{\vec{k}_2}
\hat{A}^{\gamma} _{\vec{k}_3}  \rangle $, and $\langle
\hat{A}^{\alpha} _{\vec{k}_1} \hat{A}^{\beta} _{\vec{k}_2}
\hat{A}^{\gamma} _{\vec{k}_3}  \hat{A}^{\delta} _{\vec{k}_4}\rangle
$. Here $\langle...\rangle$ means the expectation value
$\text{tr}[\hat{\rho}...]$, and $\hat{A}^{\alpha}_{\vec{k}}$ means
either $\hat{A}_{\vec{k}}$ (for $\alpha = 0$) or
$\hat{A}^{\dagger}_{\vec{k}}$ (for $\alpha = 1$).

We assume that the initial state is homogeneous. It
follows that the initial two point function can be parameterized in
terms of two functions $\mathcal{F}_1 (\vec{k})$ and $\mathcal{F}_2
(\vec{k})$ by
\bes
\label{3.1and2}
\bea
\langle \hat{A}_{\vec{k}_1} \hat{A}_{\vec{k}_2} \rangle &=& (2 \pi)^3
\delta (\vec{k}_1+\vec{k}_2) \mathcal{F}_1 (\vec{k}_1),\\
\label{3.1}
\langle  \hat{A}^{\dagger}_{-\vec{k}_1} \hat{A}_{\vec{k}_2} \rangle &=&
(2 \pi)^3 \delta (\vec{k}_1+\vec{k}_2) \mathcal{F}_2 (\vec{k}_1).
\label{3.2}
\eea
\ees
It is clear that $\mathcal{F}_2$ is real and $\mathcal{F}_1$ is
even. Under translation $\vec{x} \rightarrow \vec{x} +\vec{a} $ we
have $\hat{A}_{\vec{k}} \rightarrow \hat{A}_{\vec{k}} e^{-i
  \vec{k}. \vec{a}}$, which dictates the appearance of the specific
delta functions in Eqs.\ (\ref{3.1and2}).
For an isotropic initial state  ${\cal F}_1$ and ${\cal
  F}_2$ are functions just of $k$.

Similarly, the three point function under the assumption of
homogeneity
can be parameterized in terms of two functions $\mathcal{H}_1
(\vec{k}_1,\vec{k}_2)$ and $\mathcal{H}_2 (\vec{k}_1,\vec{k}_2)$,
defined by
\bes
\label{calHdef}
\bea
\langle \hat{A}_{\vec{k}_1} \hat{A}_{\vec{k}_2} \hat{A}_{\vec{k}_3}
\rangle  &=& (2 \pi)^3 \delta(\vec{k}_1+\vec{k}_2+\vec{k}_3)
\mathcal{H}_1 (\vec{k}_1,\vec{k}_2),\\
\langle  \hat{A}^{\dagger}_{-\vec{k}_3} \hat{A}_{\vec{k}_1}
\hat{A}_{\vec{k}_2} \rangle &=&  (2 \pi)^3
\delta(\vec{k}_1+\vec{k}_2+\vec{k}_3) \mathcal{H}_2
(\vec{k}_1,\vec{k}_2).
\eea
\ees
Both ${\cal H}_1$ and ${\cal H}_2$ are symmetric under interchange of
their arguments.  Also the other expectation values
$\langle {\hat A}_{{\vec k}_1}^\alpha {\hat A}_{{\vec k}_2}^\beta
{\hat A}_{{\vec k}_3}^\gamma \rangle$
can be expressed in terms of ${\cal H}_1$ and ${\cal H}_2$.
In the isotropic case, ${\cal H}_1$ and ${\cal H}_2$ depend only on
$k_1$, $k_2$ and on the angle between ${\vec k}_1$ and ${\vec k}_2$.
Equivalently, they are functions of $k_1$, $k_2$ and $k_3$, or of
$k$, $k_1/k_3$ and $k_2/k_3$, where $k \equiv (k_1 k_2 k_3)^{1/3}$.

Finally, the initial 4-point function can be parameterized by three
functions $\mathcal{G}_1$, $\mathcal{G}_2$, and $\mathcal{G}_3$ of 3
vectors:
\bes
\label{calGdef}
\bea
\label{3.3a}
\langle \hat{A}_{\vec{k}_1} \hat{A}_{\vec{k}_2} \hat{A}_{\vec{k}_3}
\hat{A}_{\vec{k}_4}  \rangle  &=& (2 \pi)^6
\delta(\vec{k}_1+\vec{k}_2+\vec{k}_3+\vec{k}_4) \mathcal{G}_1
(\vec{k}_1,\vec{k}_2,\vec{k}_3), \\
\label{3.3b}
\langle \hat{A}^{\dagger}_{-\vec{k}_1} \hat{A}_{\vec{k}_2}
\hat{A}_{\vec{k}_3} \hat{A}_{\vec{k}_4}  \rangle  &=& (2 \pi)^6
\delta(\vec{k}_1+\vec{k}_2+\vec{k}_3+\vec{k}_4) \mathcal{G}_2
(\vec{k}_1,\vec{k}_2,\vec{k}_3), \\
\langle \hat{A}^{\dagger}_{-\vec{k}_1}
\hat{A}^{\dagger}_{-\vec{k}_2} \hat{A}_{\vec{k}_3} \hat{A}_{\vec{k}_4}
\rangle  &=& (2 \pi)^6 \delta(\vec{k}_1+\vec{k}_2+\vec{k}_3+\vec{k}_4)
\mathcal{G}_3 (\vec{k}_1,\vec{k}_2,\vec{k}_3).
\label{eq3.7}
\eea
\ees
These functions inherit some symmetry properties from their definitions.
The function ${\cal G}_1$ is symmetric in all three of its arguments,
and in addition obeys the identity
\be
\label{ident1}
{\cal G}_1({\vec k}_1, {\vec k}_2, {\vec k}_3) =
{\cal G}_1({\vec k}_1, {\vec k}_2, -{\vec k}_1 - {\vec k}_2 - {\vec
  k}_3),
\ee
which follows from the invariance of the left hand side of Eq.\
(\ref{3.3a}) under ${\vec k}_3 \leftrightarrow {\vec k}_4$.
Similarly the function ${\cal G}_2$ is symmetric in it second and
third argument, and obeys the identities
\bes
\bea
{\cal G}_2({\vec k}_1, {\vec k}_2, {\vec k}_3) &=&
{\cal G}_2({\vec k}_1, -{\vec k}_1 - {\vec k}_2 - {\vec k}_3, {\vec k}_3), \\
{\cal G}_2({\vec k}_1, {\vec k}_2, {\vec k}_3) &=&
{\cal G}_2({\vec k}_1, {\vec k}_2, -{\vec k}_1 - {\vec k}_2 - {\vec k}_3).
\eea
\ees
The function
${\cal G}_3$ is symmetric in its first two arguments, and obeys the
identity
\be
{\cal G}_3({\vec k}_1, {\vec k}_2, {\vec k}_3) =
{\cal G}_3({\vec k}_1, {\vec k}_2,-{\vec k}_1 - {\vec k}_2 - {\vec
  k}_3).
\label{ident4}
\ee
In addition
taking the complex conjugate of Eq.\ (\ref{eq3.7}) gives the identity
\be
{\cal G}_3({\vec k}_1,{\vec k}_2, {\vec k}_3)^* =
{\cal G}_3({\vec k}_1 + {\vec k}_2 + {\vec k}_3, - {\vec k}_3, - {\vec
  k}_2).
\ee
We also decompose the four point function into its disconnected or
Gaussian piece, and its connected piece:
\begin{subequations}
\label{decom}
\begin{eqnarray}
\mathcal{G}_1(\vec{k}_1,\vec{k}_2,\vec{k}_3) &=&  \delta(\vec{k}_1+\vec{k}_2)  \mathcal{F}_1 (\vec{k}_1) \mathcal{F}_1
 (\vec{k}_3)
 +\delta(\vec{k}_1+\vec{k}_3)  \mathcal{F}_1(\vec{k}_1)\mathcal{F}_1 (\vec{k}_2)
+ \delta(\vec{k}_1+\vec{k}_4)
\mathcal{F}_1(\vec{k}_1) \mathcal{F}_1(\vec{k}_2)  \nonumber \\
&&
+\Gamma_1 (\vec{k}_1,\vec{k}_2,\vec{k}_3).
\end{eqnarray}
\begin{eqnarray}
\mathcal{G}_2(\vec{k}_1,\vec{k}_2,\vec{k}_3) &=&  \delta(\vec{k}_1+\vec{k}_2)  \mathcal{F}_2 (\vec{k}_1) \mathcal{F}_1
 (\vec{k}_3)
 +\delta(\vec{k}_1+\vec{k}_3)  \mathcal{F}_2(\vec{k}_1)\mathcal{F}_1 (\vec{k}_2)
+ \delta(\vec{k}_1+\vec{k}_4)
\mathcal{F}_2(\vec{k}_1) \mathcal{F}_1(\vec{k}_2)  \nonumber \\
&&
+\Gamma_2 (\vec{k}_1,\vec{k}_2,\vec{k}_3).
\end{eqnarray}
\begin{eqnarray}
\mathcal{G}_3(\vec{k}_1,\vec{k}_2,\vec{k}_3) &=&  \delta(\vec{k}_1+\vec{k}_2)  \mathcal{F}_1 (\vec{k}_1)^* \mathcal{F}_1
 (\vec{k}_3)
 +\delta(\vec{k}_1+\vec{k}_3)  \mathcal{F}_2(\vec{k}_1)\mathcal{F}_2 (\vec{k}_2)
+ \delta(\vec{k}_1+\vec{k}_4)
\mathcal{F}_2(\vec{k}_1) \mathcal{F}_2(\vec{k}_2)  \nonumber \\
&&
+\Gamma_3 (\vec{k}_1,\vec{k}_2,\vec{k}_3),
\end{eqnarray}
\end{subequations}
Here the functions $\Gamma_1$, $\Gamma_2$ and $\Gamma_2$ parameterize
the connected piece.

We note that Agullo and Parker \cite{Parker} chose an initial density
matrix which is diagonal on the eigenbasis of mode occupation number
$\hat{N}_{\vec{k}} = \hat{A}^{\dagger}_{\vec{k}}
\hat{A}_{\vec{k}}$ associated with the Bunch-Davies vacuum. This
choice imposes the following restrictions on
the functions $\mathcal{F}_i$ , $\mathcal{H}_{i}$ , and
$\mathcal{G}_{i}$:
\bes
\bea
\label{f1vanish}
\mathcal{F}_1 &=&0 , \\
{\cal H}_1 &=&  {\cal H}_2 = 0,\\
{\cal G}_1 &=& {\cal G}_2 = 0.
\eea
\ees
Our results for the spectrum and bispectrum below differ from theirs
by terms involving these functions which vanish for their initial state.
%
%

Now there are two different versions of the Bunch-Davies initial
vacuum state.  There is the state $\left| 0\right>$ which satisfies
${\hat A}_{\vec k} \left| 0 \right> = 0$, the ground state of the free
theory defined by the quadratic action (\ref{2.6}).  There is also the
dressed vacuum $\left| \Omega \right>$, defined as the ground
state of the Hamiltonian of the full theory including nonlinear
interaction terms given in Eq.\ (\ref{inter}) above.
The standard Bunch-Davies result for the bispectrum, first calculated
by Maldacena \cite{maldacena}, is the result for the dressed vacuum.
The functions ${\cal F}_i$, ${\cal H}_2$, ${\cal G}_i$ and $\Gamma_i$
all vanish in both vacua, $\left| 0 \right>$ and $\left| \Omega
\right>$, to leading order in the interaction (\ref{inter})
(see Appendix \ref{app:dressedvacuum}).
However the function ${\cal H}_1$ vanishes only
in the unperturbed vacuum $\left| 0 \right>$, and not in
the dressed vacuum $\left| \Omega \right>$.
We therefore define a modified version of this function by
subtracting the dressed vacuum contribution
\be
{\hat {\cal H}}_1 \equiv {\cal H}_1 - {\cal H}_1^{0,{\rm
    dr}}.
\label{hatdef}
\ee
Here ${\cal H}_1^{0,{\rm dr}}$ means the value
of ${\cal H}_1$ in the dressed vacuum $\left| \Omega \right>$,
which we compute in Appendix \ref{app:dressedvacuum}.
We will express our result below for the bispectrum in terms of
${\cal F}_i$, ${\hat {\cal H}}_1$,${\cal H}_2$,
and $\Gamma_i$, so that when these functions vanish our result
reduces to that of Maldacena \cite{maldacena}.


\subsection{Constraints from backreaction considerations}
\label{sec:back}


The energy density and pressure due to the non-vacuum initial
state must be small enough that it does not significantly perturb the
inflationary background solution.  We now review the order of
magnitude estimate of this constraint \cite{holt,shandera,agarwal}.
As long as we are considering modes which are well inside the horizon,
the stress energy density associated with the non-vacuum state
consists of an energy density $\rho_r$ and a pressure $p_r = \rho_r/3$.
These must be small compared to the background energy density $\sim M_p
^2 H^2 $ during inflation, yielding
\be
\rho_r \ll M_p ^2 H^2.
\label{3.9}
\ee

A more precise restriction on the initial state can be obtained from the
requirement that the slow roll parameters of the background expansion
are not perturbed to be larger than indicated by observations.
From the Friedmann equations we have
\be
\dot{H} = - \epsilon H^2 = \frac{-1}{2 M_{p} ^2}  (p + \rho),
\hspace{1 cm} \ddot{H} = 2 \epsilon \eta H^3 = \frac{-1}{2 M_{p} ^2}
[\dot{p} - 3 H (p+ \rho)].
\ee
The pressure and density here consist of the usual contributions from the
background inflaton field, together with the contributions
$\rho_r$ and $p_r$
from the
nonvacuum state of the perturbations.  The changes in the slow roll
parameters due to the radiation are then
\be
\Delta \epsilon \sim \rho_r / (M_p^2 H^2), \hspace{1.5 cm}
\Delta \eta \sim \rho_r / (\epsilon M_p^2 H^2).
\label{br}
\ee
Imposing that $\Delta \epsilon \lesssim \epsilon$ and $\Delta \eta \lesssim
\eta$ yields the constraints \cite{holt}
\bes
\label{3.11}
\bea
\rho_r &\lesssim& \epsilon M_p^2 H^2 \label{worse}, \\
\rho_r &\lesssim& \epsilon
 \eta M_p^2 H^2.
\label{better}
\eea
\ees
The constraints (\ref{3.11}) are more precise than
Eq. (\ref{3.9}), and
strongly restrict the number of quanta
present in the initial state. Note that the upper bound (\ref{worse}) is just the kinetic energy
$\dot{\phi}^2 \sim \epsilon M_p ^2 H^2$
of the  inflaton field.

We impose the constraints (\ref{3.11}) at the initial time $\tau_0$.
If they are satisfied then, they will be satisfied at all subsequent
times since the radiation energy density will fall off rapidly while
the inflaton energy density is approximately constant.

We can approximately characterize initial states in terms of the mean
mode occupation number $N_{\rm occ}(k)$, related to the functions
${\cal F}_1$ and ${\cal F}_2$ by \footnote{For the special case of
  generalized vacuum states, the estimate (\ref{est}) is valid for $N_{\rm occ}
  \gtrsim 1$, but must be replaced by ${\cal F}_2 \sim N_{\rm occ}$,
  ${\cal F}_1 \sim \sqrt{N_{\rm occ}}$ when $N_{\rm occ} \ll 1$, cf.\
  Eq.\ (\protect{\ref{7.13}}) below.  Because we are interested in general states we
will use the estimate (\ref{est}) in the remainder of this paper.}
\be
N_{\rm occ} \sim {\cal F}_1 \sim {\cal F}_2.
\label{est}
\ee
The energy density per logarithmic wavenumber is $\sim
N_{\rm occ}(k) (k/a)^4$.  In order to explore the consequences of the
constraints (\ref{3.11}), we specialize to a class of states for which
the mean occupation is a power law
in some interval $k_{\rm min} \le k \le k_{\rm max}$ of comoving wavenumber,
\be
N_{\rm occ}(k) \sim N_0 \left(\frac{k}{k_{\rm min}} \right)^{n-4}
\label{pl}
\ee
and vanishes outside that interval.  The corresponding energy density
is $\rho_r \sim N_0 E_{\rm max}^4 \chi^{-4} f_n(\chi)$, where $E_{\rm max} =
k_{\rm max} / a(\tau_0)$ is the maximum physical energy of occupied
modes, $\chi = k_{\rm max}/k_{\rm min}$, and
\begin{equation}
\label{fndef}
f_n(\chi) = \left\{ \begin{array}{lll}
         \ln \chi   &\ \ & \mbox{$n=0$,}  \\
         \frac{\chi^n-1}{n}
          &\ \ & \mbox{    $n \ne 0$.} \\
                \end{array}
        \right.
\end{equation}
Combining these
estimates with the constraint (\ref{better}) yields an upper bound on the
mode occupation number
\be
N_0 \lesssim \frac{ \epsilon \eta M_p^2 H^2 \chi^4}{f_n(\chi) E_{\rm max}^4}.
\label{upperbound}
\ee
We now explore the consequences of this bound under two different sets
of assumptions, a robust and conservative set of assumptions in the
next subsection, and a more speculative set in the following subsection.

\subsubsection{Constraints under conservative assumptions}

Our conservative assumptions are:
\begin{itemize}
\item At the initial time, the standard inflationary action (\ref{2.1})
  that we have assumed is valid as an effective field theory up to
  some cutoff energy scale $\Lambda$.

\item At the initial time, the occupied modes are all inside the
  horizon and below the cutoff, that is,
\be
H \le k_{\rm min}/a \le k_{\rm max}/a \le \Lambda.
\ee
This assumption is necessary for the validity of our method of
computation of the backreaction due to the occupied modes.
\end{itemize}
It follows from these assumptions that $E_{\rm max} \ge \chi H$, where
$\chi = k_{\rm max}/k_{\rm min}$.  Combining this with the upper bound
(\ref{upperbound}) and eliminating $\epsilon$ in favor of the measured
amplitude of the power spectrum $\Delta_{\cal R}$ using Eq.\ (\ref{4.3})
gives
\be
N_0 \lesssim \frac{1}{f_n(\chi)} \frac{\eta}{8 \pi^2 \Delta_{\cal
    R}^2}.
\label{dsds}
\ee
Now the observed value of the scalar spectral index, $n_s = 1
+ 2 \eta - 4 \epsilon \sim
0.96$ \cite{baumannetal}, suggests that
$\epsilon \sim \eta \sim 0.01$, absent any fine tuning.
Also the observed value of $\Delta_{\cal R}$ is $\Delta_{\cal R} \sim
3 \times 10^{-5}$ \cite{baumann}.  Inserting these estimates finally
gives the upper bound
\be
N_0 \lesssim \frac{10^5}{f_n(\chi)}.
\label{upper1}
\ee

The final upper bound (\ref{upper1}) on the mode occupation number
depends strongly on the width $\chi = k_{\rm max}/k_{\rm min}$ of the band
of occupied modes, as well as the power law index $n$.  If only a
small set of modes is occupied, $\chi \sim
1$, then the mode occupation number can be large compared to unity.
However, if the entire range of modes visible in the CMB are occupied,
so that $\chi \gtrsim 10^3$, then
the constraint depends on the power law index.
For example, for $n=0$, a scale invariant spectrum, $N_0$ can be large.
For $n=4$
which corresponds to $N_{\rm occ}(k) \sim ({\rm constant})$,
we obtain the constraint that
$N_0 \lesssim 10^{-7}$, which is quite restrictive.

Note that $N_0$ is the value of $N_{\rm occ}(k)$ at the
minimum value of $k$, $k = k_{\rm min}$, and is also the largest value
of $N_{\rm occ}(k)$ as long as $n \le 4$.  Therefore the constraint
(\ref{upper1}) is an absolute upper bound for $N_{\rm occ}(k)$ for
$n\le 4$.
It will also be useful later to have constraints on the mode
occupation number evaluated at the largest value of $k$,
\be
N_1 \equiv N_{\rm occ}(k_{\rm max}) = N_0 \chi^{n-4},
\ee
even though $N_{\rm occ}(k)$ can exceed $N_1$.  From Eq.\ (\ref{upper1}) we
find
\be
N_1 \lesssim \frac{10^5}{\chi^{4-n} f_n(\chi)} = \frac{10^5}{\chi^4}
\left[\frac{ n}{1 - \chi^{-n}}\right].
\label{N1bound}
\ee
The factor in square brackets is either of order unity or small, for
$|n| \lesssim ({\rm a\ few})$, so we obtain $N_1 \lesssim
10^5/\chi^4$.  In particular $N_1 \ll 1$ if the entire range of modes
visible in the CMB are occupied, $\chi \gtrsim 10^3$.

Another interesting quantity to constrain is the maximum bandwidth
$\chi_{\rm occ}$
over which the occupation number exceeds unity.  In other words, if
$N_{\rm occ} \ge 1$ for some range $k_1 \le k \le k_2$ of values of
$k$, what is the largest possible value of $\chi_{\rm occ} = k_2/k_1$ that is
compatible with the backreaction constraint?  We now argue that this
maximum bandwidth is $\sim 25$.

Consider first the case $n<4$, in which the maximum value of $N_{\rm
  occ}(k)$ is $N_0$ and is achieved at $k = k_{\rm min}$.  Then if
$N_0 <1$, the occupation number never exceeds unity, and we set
$\chi_{\rm occ} =1$ in this case (zero bandwidth).  If $N_0 >1$ the
bandwidth is given by, from Eq.\ (\ref{pl}) $1 = N_0 \chi_{\rm
  occ}^{n-4}$, and combining these gives $\chi_{\rm occ} = {\rm
  max}(1, N_0^{1/(4-n)})$.  However clearly $\chi_{\rm occ}$ cannot
exceed $\chi = k_{\rm max}/k_{\rm min}$, the bandwidth in which we
have assumed $N_{\rm occ}$ is nonzero, so we obtain
\be
\chi_{\rm occ} = {\rm min}\left\{ \chi, {\rm max}(1, N_0^{1/(4-n)}) \right\}.
\ee
A similar analysis for $n>4$ yields
\be
\chi_{\rm occ} = {\rm min}\left\{ \chi, {\rm max}(1, \chi N_0^{1/(n-4)}) \right\}.
\ee
Combining these results and using the upper bound (\ref{dsds}) for
$N_0$ gives
\be
\chi_{\rm occ} \le {\rm min}\left\{ \chi, {\rm max}\left[1, \chi^{\Theta(n-4)}
\left(
\frac{G n}{\chi^n -1} \right)^{1/|n-4|} \right]
\right\},
\label{bandwidthbound}
\ee
where $\Theta$ is the step function and $G = \eta / (8 \pi^2
\Delta_{\cal R}^2) \sim 10^5$.  We now maximize over all
values of $n$ and over values of $\chi \ge 1$.  The function of $n$ and $\chi$ on the
right hand side of Eq.\ (\ref{bandwidthbound}) is maximized at $n =4$
and at $\chi = (1 + 4 G)^{1/4}$, and the maximum value is
$(1 + 4 G)^{1/4}$.  Thus we obtain
\be
\chi_{\rm occ} \lesssim (1 + 4 G)^{1/4} \sim 25.
\label{bandwdithbound1}
\ee

Finally we note that the upper limit (\ref{bandwdithbound1}) on the
bandwidth of occupied modes is weakened if one uses the less stringent
backreaction constraint (\ref{worse}) instead of the more stringent
constraint (\ref{better}).
This yields
\be
\chi_{\rm occ} \lesssim (1 + 4 G/\eta)^{1/4} \sim 100,
\label{bandwdithbound2}
\ee
which was the upper limit obtained in Ref. \citep{Flauger:2013hra}.

\subsubsection{Constraints under more speculative assumptions}

Next, an upper bound on $N_0$ which is somewhat stronger than the
bound (\ref{upper1})
can be obtained if one assumes that
the backreaction is small not just at the chosen initial time, but also at
all previous times, continuing into the past until the energies of the
occupied modes are blueshifted up to the cutoff scale $\Lambda$.
This assumption is not particularly well motivated, since it is
possible, for example, for a multifield model to behave like a single
field model in certain regimes, and going backwards in time a
transition from single field behavior to to multifield behavior can occur
while $k_{\rm   max}/a(\tau) \ll \Lambda$.  Nevertheless, if one makes
this assumption, then from Eq.\ (\ref{upperbound}) evaluated at $E_{\rm max}
\sim \Lambda$ we obtain
\be
N_0 \lesssim \frac{ \epsilon \eta M_p^2 H^2 \chi^4}{f_n(\chi) \Lambda^4}.
\label{upperbound2}
\ee
Next, we demand that the background inflationary solution be within
the domain of the effective field theory.  The action (\ref{2.1})
will have correction terms such as $(\nabla \phi)^4/\Lambda^4$ that
are suppressed by powers of the cutoff.  This correction term must
be small compared to $(\nabla \phi)^2$ when evaluated on the
background solution, which yields the condition
$\epsilon M_p^2 H^2 \ll \Lambda^4$.
Combining this with Eq.\ (\ref{upperbound2}) yields
\begin{equation}
\label{fndef1}
N_0 \le \eta \frac{\chi^4}{f_n(\chi)}
= \left\{ \begin{array}{lll}
         \frac{\eta\,  \chi^4}{\ln \chi}   &\ \ & \mbox{$n=0$,} \\
         \frac{\eta \, n \, \chi^4}{\chi^n-1}
          &\ \ & \mbox{    $n \ne 0$.} \\
                \end{array}
        \right.
\end{equation}
Thus the mode occupation number must be small
compared to unity if $\chi \sim 1$, assuming $\eta \sim 0.01$.
It can be large however for $\chi \gg 1$ if $n<4$.

\section{Scalar power spectrum }
\label{sec:power}

The scalar power spectrum $\mathcal{P}_{\mathcal{R}}$ of the perturbations is
is defined in terms of  the
momentum space equal time  two-point function of the scalar curvature
perturbation  $\hat{\mathcal{R}}_{\vec{k}}(\tau)$:
\be
\langle \hat{\mathcal{R}}_{\vec{k}_1}(\tau)
\hat{\mathcal{R}}_{\vec{k}_2}(\tau)  \rangle =  (2 \pi)^3
\delta(\vec{k}_1+ \vec{k}_2 ) \mathcal{P} _{\mathcal{R}}  ({\vec
  k}_1).
\label{4.1}
\ee
Here it is assumed that the conformal time $\tau$ is
chosen to be any time after the modes have exited the horizon,
so that the right hand side is independent of $\tau$.
The dimensionless power spectrum $\Delta_{\cal R}$ is then defined by
\be
 \mathcal{P}_{\mathcal{R}}({\vec k}) \equiv \frac{2 \pi ^2} {k^3}
 \Delta^2 _\mathcal{R} ({\vec k}).
\ee
The scalar power spectrum has two contributions,
\be
\mathcal{P}_{\mathcal{R}}({\vec k}) =
\mathcal{P}^{(0)}_{\mathcal{R}}({\vec k})
+ \mathcal{P}^{(1)}_{\mathcal{R}}({\vec k}),
\ee
a leading order term $\mathcal{P}^{(0)}_{\mathcal{R}}({\vec k})$
neglecting the effect of interactions, and a
subleading term $\mathcal{P}^{(1)}_{\mathcal{R}}({\vec k})$
arising from the interactions, cf.\ Sec.\
\ref{sec:interactions} above.
The subleading term is smaller by a factor $\sim
H \sqrt{\epsilon}/M_p$.
We will consider
both contributions.

\subsection{Leading order power spectrum}

For the Bunch-Davies vacuum initial state, this power spectrum is
given by \cite{baumann}
\be
\mathcal{P}_{{\cal R},0}^{(0)}(k) = | \mathcal{R}(k, \tau_k)|^2,
\label{p0}
\ee
where the
subscript $0$ is used to indicate that this quantity is calculated
in the Bunch-Davies vacuum, $\mathcal{R}(k,\tau)$ is the mode
function defined by the condition (\ref{2.13}), and $\tau_k$ is the
time of horizon crossing given by $k = a(\tau_k) H$. Thus
\be
\Delta^2_{\mathcal{R},0}(k)=  \frac{1}{2 M_p ^2 \epsilon(\tau_k)} \left[
\frac{H(\tau_k)}{2 \pi} \right]^2
\label{4.3}
\ee
Henceforth we shall denote the mode functions by $\mathcal{R} (k)$
only and restore the time dependence only when necessary.

For a general homogeneous initial state, we compute the leading order power spectrum
by inserting the mode expansion (\ref{2.9}) into the power spectrum
definition (\ref{4.1}), and simplifying using the definitions
(\ref{3.1and2}) of the two point functions ${\cal F}_1$
and ${\cal F}_2$.  This yields\footnote{An additional contribution to the change in
the measured spectral index is the backreaction effect
(\protect{\ref{br}}).  Observations constrain the sum of the
backreaction contribution and the direct contribution
(\protect{\ref{ps0}}).  One could imagine evading the backreaction
constraints (\protect{\ref{3.11}}) via a cancellation between these
two contributions, but this would require a fine tuning.}
 (see, e.g., Ref.\ \cite{2005PhRvD..72d3515A})
\be
\mathcal{P}^{(0)}_{\mathcal{R}} (\vec{k}) =  \mathcal{P}^{(0)}_{\mathcal{R},0}
(\vec{k}) \big[ 1+ \mathcal{F}_2 (-\vec{k})  +  \mathcal{F}_2 (\vec{k})  -
2 \text{Re} \{\mathcal{F}_1 (\vec{k}) \}  \big ],
\label{ps0}
\ee
where Re means the real part of.
Here we used the fact that $\mathcal{R}(k,\tau)$ is
purely imaginary at late times according to Eq.\ (\ref{2.13}).  The
terms in the square brackets after the $1$ indicate the effect
of the presence of quanta in the initial state.  The result
(\ref{ps0}) agrees with that of Agullo and Parker \cite{Parker},
except that they omit the final term in the square bracket,
which vanishes for their class of initial states, cf.\ Eq.\
(\ref{f1vanish}) above.

The leading order scalar power spectrum (\ref{ps0}) has been calculated for
variety of different choices of initial states \cite{Parker, kundu,
ashoorion, dey}. For some specific initial states
the deviation from the Bunch-Davies vacuum case vanishes
\cite{ashoorion,kundu}.
Current observational constraints on the flatness of the power
spectrum constrain the initial state, ruling out models where
the scale invariance of the power spectrum is
strongly violated, for example Ref.\ \cite{dey}.

\subsection{Subleading contributions to power spectrum due to interactions}
\label{sec:subleadingpower}

To calculate the power spectrum (\ref{4.1}) to subleading order we use
time-dependent perturbation theory.
We choose the origin of conformal time
so that the value of $\tau$ on the left hand side of Eq.\ (\ref{4.1})
is $\tau=0$, that is, $\tau =0$ occurs after the modes exit the horizon.
At leading order in perturbation theory we have
\bea
 \langle  \hat{\mathcal{R}}_{\vec{k}_1}(0) \hat{ \mathcal{R}}_{\vec{k}_2}(0)  \rangle
&=& \langle \hat{\mathcal{R}}^{I}_{\vec{k}_1}(0)
\hat{\mathcal{R}}^{I}_{\vec{k}_2}(0) \rangle
 - i \int_{\tau_0}^ 0
d\tau \left< \left[  \hat{\mathcal{R}}^{I}_{\vec{k}_1}(0)
\hat{\mathcal{R}}^{I}_{\vec{k}_2}(0), {\hat H}_{\rm int}^I (\tau)\right] \right>
 + O ({\hat H}_{\rm int}^2).
\label{5.4a}
\eea
Here the superscripts $I$ denote interaction picture operators,
defined with respect to the initial conformal time $\tau_0$ at which
we specify the initial state, and
${\hat H}_{\rm int}$ is the interaction Hamiltonian given by Eq.\
(\ref{inter}).
We now rewrite the field operators $\hat{\mathcal{R}}$ on the right hand side
in terms of the redefined field operator
$\hat{\mathcal{R}}_c$ using
Eq.\ (\ref{5.2}).
We then insert the mode expansion (\ref{2.9a}) for the redefined interaction picture operators
together with the explicit expression (\ref{inter}) for the
interaction Hamiltonian, and simplify using the notations introduced
in Sec.\ \ref{sec:notation}.  We neglect any evolution in the slow
roll parameters that occurs between the horizon exits of the different
modes, that is, we treat these slow roll parameters as constants.  The
calculation is analogous to the calculation of the bispectrum detailed in
Appendix \ref{app:bispectrum}.

The final result for the subleading power spectrum is
\be \mathcal{P}^{(1)}_{\mathcal{R}}(\vec{k}) =- \frac{ i H^3 (
  \frac{3}{2} \epsilon- \eta)  }{16 M_p ^3 k ^3 \epsilon
  ^{\frac{3}{2}}} \Pi_1 (\vec{k}) - \frac{ i H^3   }{16 M_p ^3 k
  ^3 \epsilon ^{\frac{1}{2}}} \Pi_2 (\vec{k})+ \frac{H^3}{8 \epsilon
  ^{\frac{1}{2}} M_p ^3 k ^3} \Pi_3 (\vec{k}).
\label{subleading}
\ee
Here the functions $\Pi_1$, $\Pi_2$ and $\Pi_3$ are dimensionless
functions of momentum.  The function $\Pi_1$ is given by
\bea
\Pi_1 (\vec{k}) &=& \tilde{\Phi}_1 (\vec{k})- \tilde{\Phi}_{2,1}
(\vec{k})- \tilde{\Phi}_{2,2} (\vec{k}) -\tilde{\Phi}_{2,3}
(\vec{k})
+\tilde{\Phi}_{2,3}(-\vec{k})^*+ \tilde{\Phi}_{2,2}(-\vec{k})^*
\nonumber \\ &&
+
\tilde{\Phi}_{2,1}(-\vec{k})^* - \tilde{\Phi}_1(-\vec{k})^* + (\vec{k} \rightarrow -\vec{k}),
\eea
where
\be
\tilde{\Phi}_1 (\vec{k}) = \int d^3p \hspace{0.5mm} p^{-3/2} q^{-3/2}
k^{3/2} \hat{\mathcal{H}}_1(\vec{k},\vec{p}),
\label{eee1}
\ee
\be
\tilde{\Phi}_{2,1} (\vec{k}) = \int d^3p \hspace{0.5mm} p^{-3/2} q^{-3/2} k^{3/2} \mathcal{H}_2(\vec{k},\vec{p}), \ee
\be
\tilde{\Phi}_{2,2} (\vec{k}) = \int d^3p\hspace{0.5mm} p^{-3/2} q^{-3/2} k^{3/2} \mathcal{H}_2(\vec{k},\vec{q}), \ee
\be
\tilde{\Phi}_{2,3} (\vec{k}) = \int d^3p\hspace{0.5mm} p^{-3/2}
q^{-3/2} k^{3/2} \mathcal{H}_2(\vec{p},\vec{q}),
\label{eee4}
\ee
and $\vec{q} = -\vec{k}-\vec{p}$.  Similarly the function
$\Pi_2$ is given by
\bea
\Pi_2(\vec{k}) &=& \bar{\Phi}_1(\vec{k})- \bar{\Phi}_{2,1}(\vec{k})
- \bar{\Phi}_{2,2}(\vec{k}) -\bar{\Phi}_{2,3}(\vec{k})
 +\bar{\Phi}_{2,3}(-\vec{k})^*+
\bar{\Phi}_{2,2}(-\vec{k})^*
\nonumber \\ &&
+ \bar{\Phi}_{2,1}(-\vec{k}_1)^* - \bar{\Phi}_1(-\vec{k}_1)^*
 + (\vec{k} \rightarrow -\vec{k}),
\eea
where the functions $\bar{\Phi}$ are defined
by equations analogous to Eqs.\ (\ref{eee1}) - (\ref{eee4})
but with the factor $p^{-3/2} q^{-3/2} k^{3/2} $ in the integrand
replaced with $p^{-3/2} q^{1/2} k^{-1/2}$.  Lastly, the function
$\Pi_3$
in Eq.\ (\ref{subleading})
is given by
\begin{eqnarray}
\Pi_3(\vec{k})&=&    \int d^3 p \sqrt{p s k} \Big[ \frac{1}{k ^2}
+ \frac{1}{p^2} +   \frac{1}{s^2} \Big ]
\nonumber\\ &&
\times \Bigg \{  h(p,s) \Big [\hat{{\cal H}}_1(\vec{p},\vec{s}) -
{\cal H}_2(\vec{p},\vec{s}) \Big] - h(-p,s)  \Big [{\cal
  H}_2(-\vec{k},\vec{p}) - {\cal H}_2(-\vec{s},\vec{k})^* \Big]
\nonumber\\
&&-  h(p,-s) \Big [{\cal H}_2(-\vec{k},-\vec{s}) - {\cal
  H}_2(-\vec{p},\vec{k})^* \Big] + h(-p,-s)  \Big [{\cal
  H}_2(-\vec{p},-\vec{s})^*  - \hat{{\cal
    H}}_1(-\vec{p},\vec{k})^* \Big] \Bigg \} \nonumber\\
&&+ (\vec{k} \rightarrow -\vec{k}) ,
\end{eqnarray}
where ${\vec s} = {\vec k} - {\vec p}$.  Here
the function $h(a,b)$ is
\bea
h(a,b) &=& \int_0^{\tau_0} d\tau \left[ e^{-i \tau(a + b + k)} - e^{-i
    \tau(a + b - k)} \right] \nonumber \\
&=&\frac{i}{a + b - k} \left[ 1 - e^{-i \tau_0 (a + b - k)} \right] -
  \frac{i}{a + b + k} \left[ 1 - e^{-i \tau_0 (a + b + k)} \right].
\eea

We note that the subleading contribution (\ref{subleading})
to the power spectrum depends only on the three point function of the
initial state, parameterized by the functions $\mathcal{H}_i$, and so
it vanishes for Gaussian initial states. This includes the
Bunch-Davies vacuum state.

\section{Primordial non-Gaussianities: bispectrum of scalar perturbations}
\label{sec:bispectrum}

To predict the non-Gaussianity of the primordial fluctuations one
needs to consider higher order correlation functions. The full
computation of the momentum space three point correlation function of the
comoving curvature perturbation field for a vacuum initial state in a
single field model was done by Maldacena \cite{maldacena}. Here we
calculate the three point function for the class of non-vacuum homogeneous
initial states described in section \ref{sec:initial}.  This calculation
is a generalization of the one done by Agullo and Parker \cite{Parker}.

The three point function of the comoving curvature perturbation field  $
\hat{\mathcal{R}}_{\vec{k}}(\tau)$ is parameterized as
\be
\langle\hat{\mathcal{R}}_{\vec{k}_1}(\tau)\hat{\mathcal{R}}_{\vec{k}_2}(\tau)\hat{\mathcal{R}}_{\vec{k}_3}(\tau)
\rangle = (2 \pi)^3 \delta(\vec{k}_1+\vec{k}_2+\vec{k}_3)
\mathcal{B}(\vec{k}_1,\vec{k}_2,\vec{k}_3),
\label{calBdef}
\ee
where $ \mathcal{B}(\vec{k}_1,\vec{k}_2,\vec{k}_3)$ is called the
bispectrum.  Here it is assumed that the conformal time $\tau$ is
chosen to be any time after all three modes have exited the horizon,
so that the right hand side is independent of $\tau$.
We now fix values of ${\vec k}_1$, ${\vec k}_2$, ${\vec k}_3$, which we
assume to be nonzero, and we choose the origin of conformal time as before
so that the value of $\tau$ on the left hand side of Eq.\ (\ref{calBdef})
is $\tau=0$, that is, $\tau =0$ occurs after the modes exit the horizon.

To calculate the correlation function (\ref{calBdef}) we use
time-dependent perturbation theory as in Sec.\ \ref{sec:subleadingpower} above.
At leading order in perturbation theory we have
\bea
 \langle  \hat{\mathcal{R}}_{\vec{k}_1}(0) \hat{ \mathcal{R}}_{\vec{k}_2}(0) \hat{\mathcal{R}}_{\vec{k}_3} (0) \rangle
&=& \langle \hat{\mathcal{R}}^{I}_{\vec{k}_1}(0)
\hat{\mathcal{R}}^{I}_{\vec{k}_2}(0)
\hat{\mathcal{R}}^{I}_{\vec{k}_3} (0) \rangle
\nonumber \\
&& - i \int_{\tau_0}^ 0
d\tau \left< \left[  \hat{\mathcal{R}}^{I}_{\vec{k}_1}(0)
\hat{\mathcal{R}}^{I}_{\vec{k}_2}(0)
\hat{\mathcal{R}}^{I}_{\vec{k}_3} (0), {\hat H}_{\rm int}^I (\tau)\right] \right>
 + O ({\hat H}_{\rm int}^2).
\label{5.4}
\eea
Here the superscripts $I$ denote interaction picture operators,
${\hat H}_{\rm int}$ is the interaction Hamiltonian given by Eq.\
(\ref{inter}), and $\tau_0$ is the conformal time at which we specify
the initial state.
We now rewrite the field operators $\hat{\mathcal{R}}$ on the right
hand side in terms of the
redefined field operator $\hat{\mathcal{R}}_c$ using
Eq.\ (\ref{5.2}).
We then insert the mode expansion (\ref{2.9a}) for the redefined interaction picture operators
together with the explicit expression (\ref{inter}) for the
interaction Hamiltonian, and simplify using the notations introduced
in Sec.\ \ref{sec:notation}.  We neglect any evolution in the slow
roll parameters that occurs between the horizon exits of the different
modes, that is, we treat these slow roll parameters as constants.  The
details of this calculation are given in Appendix
\ref{app:bispectrum}.

We note that the interaction Hamiltonian (\ref{inter}) is cubic in the
fields, so the second term in Eq.\ (\ref{5.4}) contains six factors of
field operators.  However, these factors occur inside a commutator,
and when this commutator is evaluated explicitly the number of factors
of field operators is reduced to four.  Therefore we only require the
two, three and four point functions of the initial state, and not any
higher point functions.

The final result for the bispectrum can be written as follows:
\be
{\cal B}({\vec k}_1, {\vec k}_2, {\vec k}_3) =
{\cal B}_{0,{\rm dr}}({\vec k}_1, {\vec k}_2, {\vec k}_3) +
{\cal B}_{II}({\vec k}_1, {\vec k}_2, {\vec k}_3) +
{\cal B}_{III}({\vec k}_1, {\vec k}_2, {\vec k}_3) +
{\cal B}_{IV}({\vec k}_1, {\vec k}_2, {\vec k}_3).
\label{finalresult}
\ee
Here ${\cal B}_{0,{\rm dr}}$ is the bispectrum for the initial,
dressed, Bunch-Davies vacuum, as calculated by Maldacena \cite{maldacena}.
The remaining terms are corrections due to the non-vacuum initial state.
The term ${\cal B}_{II}$ is determined by the two point function of the
initial state, ${\cal B}_{III}$ by the three point function, and
${\cal B}_{IV}$ by the connected part of the four point function.
We now discuss these various contributions in turn.

\subsection{Vacuum contribution}

The vacuum contribution is \cite{maldacena}
\be
\label{vac}
{\cal B}_{0,{\rm dr}}({\vec k}_1, {\vec k}_2, {\vec k}_3)
 = \frac{H^4}{16 M_p^4 \epsilon^2 k_1^3 k_2^3 k_3^3}
\left[ \frac{1}{2} \epsilon \sigma_1 \sigma_2 - (\eta - \epsilon) \sigma_3 + 2 \epsilon (\sigma_2^2 - \sigma_4)/\sigma_1 \right],
\ee
where we have defined
\be
\sigma_p \equiv \sum_{i=1}^3 k_i^p,
\ee
for $p = 1,2,3,4$.  This is the result that applies when ${\cal F}_i =
{\hat {\cal H}}_1 = {\cal H}_2  = {\Gamma}_i = 0$.

\subsection{Contribution from two point function}

The piece ${\cal B}_{II}$ of the bispectrum depends only on the
functions ${\cal F}_1({\vec k})$ and ${\cal F}_2({\vec
  k})$ that parameterize the initial two point function,
given by Eqs.\ (\ref{3.1and2}) and (\ref{hatdef}) above.
For a
Gaussian initial state, as studied previously in Refs.\
\cite{porrati1,porrati2,holt,meer,meer1,ganc,shandera,kundu,das}, this
gives the entire bispectrum.
The result is
\bea
\label{calBII}
{\cal B}_{II}({\vec k}_1, {\vec k}_2, {\vec k}_3) &=&
\frac{H^4 (\frac{3}{2}
  \epsilon - \eta)}{16 M_p^4 \epsilon^2 k_1^2 k_2^2 k_3^2}
  {\cal A}_{II}({\vec k}_1, {\vec k}_2, {\vec k}_3) +
\frac{H^4}{32 M_p^4 \epsilon k_1^2 k_2^2 k_3^2}
  {\tilde {\cal A}}_{II}({\vec k}_1, {\vec k}_2, {\vec k}_3)
\nonumber \\
&&- \frac{i H^4}{8 M_p^4 \epsilon k_1^2 k_2^2 k_3^2}  {\hat {\cal A}}_{II}({\vec k}_1, {\vec k}_2, {\vec k}_3).
\eea
Here the dimensionless amplitude ${\cal A}_{II}$ is given by
\be
{\cal A}_{II} = \frac{k_3^2}{k_1 k_2} \left[ {\bar {\cal F}}({\vec
    k}_1) {\bar {\cal F}}({\vec k}_2)
+ {\bar {\cal F}}({\vec k}_1)
+ {\bar {\cal F}}({\vec k}_2)
\right] +
{\rm cyclic\ perms},
\ee
where ''cyclic perms'' means two terms obtained from the original term
by cyclicly permuting ${\vec k}_1, {\vec k}_2$ and ${\vec k}_3$.
Also we have defined the function
\be
{\bar {\cal F}}({\vec k}) \equiv  {\cal F}_2({\vec k}) +
{\cal F}_2(-{\vec k}) - 2 {\rm Re} \, {\cal F}_1({\vec k}),
\ee
which is same combination of ${\cal F}_1$ and ${\cal F}_2$ that appears
in the power spectrum (\ref{ps0}).
The dimensionless amplitude ${\tilde {\cal A}}_{II}$ is given by
\be
{\tilde {\cal A}}_{II} = \frac{k_1^2+ k_2^2}{k_1 k_2} \left[  {\bar {\cal
      F}}({\vec
    k}_1) {\bar {\cal F}}({\vec k}_2) + {\bar {\cal F}}({\vec k}_1)
+ {\bar {\cal F}}({\vec k}_2)
 \right] +
{\rm cyclic\ perms}.
\ee
The dimensionless amplitude ${\hat {\cal A}}_{II}$ is given by
\begin{eqnarray}
{\hat {\cal A}}_{II} = &&k_1 k_2 k_3 \sum_{i=1}^{3} \frac{1}{k_i ^2}
\Bigg (  r(-k_t,{\tilde k}_3)  \bigg[ {\cal F}_2 (\vec{k}_1) {\cal F}_2
(\vec{k}_2)+   {\cal F}_1 (\vec{k}_1) {\cal F}_1 (\vec{k}_2) \bigg]
\nonumber\\
&& +\Bigg \{  r(-{\tilde k}_2, \tilde{k}_1) \bigg[ {\cal F}_2 (-\vec{k}_1)
{\cal F}_1 (\vec{k}_2) + {\cal F}_2
(\vec{k}_2) {\cal F}^*_1 (\vec{k}_1) \bigg] + r(-k_t, {\tilde k}_3){\cal F}_2
(\vec{k}_2) {\cal F}_1 (\vec{k}_1)   \nonumber \\
&&+  r({\tilde k}_1,-{\tilde k}_2) \bigg[ {\cal F}_1 (\vec{k}_2) {\cal F}_1
(\vec{k}_1)^* +  {\cal F}_2 (\vec{k}_2) {\cal F}_2
(-\vec{k}_1) \bigg]  +r(k_t, -\tilde{k}_3) {\cal F}_2 (-\vec{k}_1)
{\cal F}_1 (\vec{k}_2)^*+ (\vec{k}_1 \leftrightarrow \vec{k}_2)\Bigg \} 
  \nonumber\\ &&
+  r(-{\tilde k}_3,k_t) \bigg[ {\cal F}_2 (-\vec{k}_1) {\cal F}_2 (-\vec{k}_2) +{\cal F}_1 (\vec{k}_1)^* {\cal F}_1 (\vec{k}_2)^* \bigg]
\nonumber \\
&& + r(-k_t,-{\tilde k}_3) {\cal F}_2({\vec k}_3) + r({\tilde k}_3,k_t) {\cal F}_2(-{\vec k}_3)+  r(-{\tilde k}_3,-k_t) {\cal F}_1({\vec k}_3) +  r(k_t,\tilde{k}_3) {\cal F}_1({\vec k}_3)^*
\Bigg )
\nonumber\\
&& + {\rm cyclic\ perms},
\label{hatcalAII}
\end{eqnarray}
where
\be
k_t \equiv k_1 +k_2 + k_3, \ \ \ \ \ \ \tilde{k_i} \equiv k_t - 2 k_i.
\label{tildekdef}
\ee
Here
we have defined the function $r(a,b)$ by
\bea
r(a,b) &=& \int_{\tau_0}^0 d\tau \left[ e^{i \tau a} - e^{i \tau b} \right] \nonumber \\
&=& {1 \over ia} \left[ 1 - e^{i \tau_0 a} \right]-
 {1 \over ib} \left[ 1 - e^{i \tau_0 b} \right].
\label{rdef}
\eea

\subsection{Contribution from three point function}

The piece ${\cal B}_{III}$ of the bispectrum is
\be
\label{calBIII}
{\cal B}_{III}({\vec k}_1, {\vec k}_2, {\vec k}_3) = - \frac{ i H^3
  }{8 M_p^3 \epsilon^{3/2} k_1^2 k_2^2 k_3^2} {\cal A}_{III}({\vec k}_1,
  {\vec k}_2, {\vec k}_3),
\ee
where the dimensionless amplitude ${\cal A}_{III}$ is given by
\bea
\label{5.7}
{\cal A}_{III} &=&
(k_1 k_2 k_3)^{1/2} \bigg[
\hat{\mathcal{H}}_1 (\vec{k}_1,\vec{k}_2) - \hat{\mathcal{H}}_1 (-\vec{k}_1,-\vec{k}_2)^*-\mathcal{H}_2 (\vec{k}_1,\vec{k}_2) +\mathcal{H}_2 (-\vec{k}_1,-\vec{k}_2)^*
 -\mathcal{H}_2 (\vec{k}_1,\vec{k}_3) \nonumber \\
&& +\mathcal{H}_2 (-\vec{k}_1,-\vec{k}_3)^*
  -\mathcal{H}_2 (\vec{k}_2,\vec{k}_3) +\mathcal{H}_2
(-\vec{k}_2,-\vec{k}_3)^* \bigg],
\eea
where the functions $\hat{\mathcal{H}}_1$ and $\mathcal{H}_2$ are defined in Eqs.\
(\ref{calHdef}) and (\ref{hatdef}) above.

\subsection{Contribution from connected piece of four point function}

This piece of the bispectrum is determined by the functions ${
  \Gamma}_i({\vec k}_1, {\vec k}_2, {\vec k}_3)$ for $i = 1,2,3$ that
parameterize the connected piece of the four point function of the
initial state, defined by Eqs.\ (\ref{calGdef}), (\ref{decom}) and
(\ref{hatdef}) above.
In order to give the result we first define some notations.  We define
the functions $\Theta_{i,j}({\vec k}_1, {\vec k}_2)$ for $1 \le i \le
3$ and $1 \le j \le 4$ by
\bes
\bea
\Theta_{i,1}({\vec k}_1,{\vec k}_2) &=& \int d^3 p \, p^{-3/2} q^{1/2}
{\Gamma}_i(\vec{k}_1,\vec{k}_2,\vec{p}),\\
\Theta_{i,2}({\vec k}_1,{\vec k}_2) &=& \int d^3 p \, p^{-3/2} q^{1/2}
{\Gamma}_i(\vec{p}, \vec{k}_1,\vec{k}_2),\\
\Theta_{i,3}({\vec k}_1,{\vec k}_2) &=& \int d^3 p \, p^{-3/2} q^{1/2}
{\Gamma}_i(\vec{q}, \vec{k}_1,\vec{k}_2),\\
\Theta_{i,4}({\vec k}_1,{\vec k}_2) &=& \int d^3 p \, p^{-3/2} q^{1/2}
{\Gamma}_i(\vec{p},\vec{q}, \vec{k}_1).
\eea
\ees
Here on the right hand sides $\vec{k}_3 = - {\vec k}_1 - {\vec k}_2$
and $\vec{q} = {\vec k}_3 - {\vec p}$.
We also define barred versions of some of these functions by modifying
the weighting factor in the integrand:
\bes
\bea
{\bar \Theta}_{i,1}({\vec k}_1,{\vec k}_2) &=& \int d^3 p \, p^{-3/2} q^{-3/2}
{\Gamma}_i(\vec{k}_1,\vec{k}_2,\vec{p}),\\
{\bar \Theta}_{i,2}({\vec k}_1,{\vec k}_2) &=& \int d^3 p \, p^{-3/2} q^{-3/2}
{\Gamma}_i(\vec{p}, \vec{k}_1,\vec{k}_2),\\
{\bar \Theta}_{i,4}({\vec k}_1,{\vec k}_2) &=& \int d^3 p \, p^{-3/2} q^{-3/2}
{\Gamma}_i(\vec{p},\vec{q}, \vec{k}_1).
\eea
\ees
These functions are not all independent; it follows from the
identities (\ref{ident1}) -- (\ref{ident4}) that
${\bar \Theta}_{1,1} = {\bar \Theta}_{1,2} = {\bar \Theta}_{1,4}$,
${\bar \Theta}_{2,2} = {\bar \Theta}_{2,4}$
$\Theta_{1,1} = \Theta_{1,2} = \Theta_{1,3} = \Theta_{1,4}$,
and $\Theta_{2,2} = \Theta_{2,4}$.  Also the functions
${\bar \Theta}_{1,1}$, ${\bar \Theta}_{2,2}$, ${\bar \Theta}_{3,1}$,
${\bar \Theta}_{3,4}$, $\Theta_{1,1}$, $\Theta_{2,2}$, $\Theta_{2,3}$,
$\Theta_{3,1}$ and $\Theta_{3,4}$ are all symmetric under the
interchange of ${\vec k}_1$ and ${\vec k}_2$.

The result for the bispectrum is
\bea
\label{calBIV}
{\cal B}_{IV}({\vec k}_1, {\vec k}_2, {\vec k}_3) &=&
\frac{H^4 (\frac{3}{2}
  \epsilon - \eta)}{32 M_p^4 \epsilon^2 k_1^2 k_2^2 k_3^2}
  {\cal A}_{IV}({\vec k}_1, {\vec k}_2, {\vec k}_3) +
\frac{H^4}{32 M_p^4 \epsilon k_1^2 k_2^2 k_3^2}
  {\tilde {\cal A}}_{IV}({\vec k}_1, {\vec k}_2, {\vec k}_3)
\nonumber \\
&&+ \frac{i H^4}{16 M_p^4 \epsilon k_1^2 k_2^2 k_3^2}  {\hat {\cal A}}_{IV}({\vec k}_1, {\vec k}_2, {\vec k}_3).
\eea
Here the dimensionless amplitude ${\cal A}_{IV}$
is given by
\bea
\label{CalAIV}
{\cal A}_{IV} &=& k_3^2 \sqrt{k_1 k_2} \bigg[
{\bar \Theta}_{1,1}(\vec{k}_1,\vec{k}_2)+{\bar
    \Theta}_{1,1}(-\vec{k}_1,-\vec{k}_2)^*
-2{\bar \Theta}_{2,2}(\vec{k}_1,\vec{k}_2)
-{\bar \Theta}_{2,1}(\vec{k}_2,\vec{k}_1)
\nonumber \\
&&
- {\bar \Theta}_{2,1}(-\vec{k}_1,-\vec{k}_2)^*
-{\bar \Theta}_{2,1}(\vec{k}_1,\vec{k}_2)-{\bar
  \Theta}_{2,1}(-\vec{k}_2,-\vec{k}_1)^*
- 2 {\bar \Theta}_{2,2}(-\vec{k}_1,-\vec{k}_2)^*
+{\bar \Theta}_{3,4}(\vec{k}_1,\vec{k}_2)
\nonumber \\
&& +2 {\bar \Theta}_{3,2}(\vec{k}_2,\vec{k}_1)
+2 {\bar \Theta}_{3,2}(\vec{k}_1,\vec{k}_2)
+{\bar \Theta}_{3,1}(\vec{k}_1,\vec{k}_2)
\bigg] \ \ + {\rm cyclic\ perms}.
\eea
The dimensionless amplitude ${\tilde {\cal A}}_{IV}$ is given by
\bea
{\tilde {\cal A}}_{IV} &=& \sqrt{k_1 k_2} \bigg[
{\Theta}_{1,1}(\vec{k}_1,\vec{k}_2)+{
    \Theta}_{1,1}(-\vec{k}_1,-\vec{k}_2)^*
-{ \Theta}_{2,2}(\vec{k}_1,\vec{k}_2)
-{ \Theta}_{2,3}(\vec{k}_1,\vec{k}_2)
-{ \Theta}_{2,1}(\vec{k}_2,\vec{k}_1)
\nonumber \\
&&
- { \Theta}_{2,1}(-\vec{k}_1,-\vec{k}_2)^*
-{ \Theta}_{2,1}(\vec{k}_1,\vec{k}_2)-{
  \Theta}_{2,1}(-\vec{k}_2,-\vec{k}_1)^*
-  {\Theta}_{2,2}(-\vec{k}_1,-\vec{k}_2)^*
\nonumber \\
&& -  {\Theta}_{2,3}(-\vec{k}_1,-\vec{k}_2)^*
+{\Theta}_{3,4}(\vec{k}_1,\vec{k}_2)
+ {\Theta}_{3,2}(\vec{k}_2,\vec{k}_1)
+ {\Theta}_{3,3}(\vec{k}_2,\vec{k}_1)
+ {\Theta}_{3,2}(\vec{k}_1,\vec{k}_2)
\nonumber \\ &&
+ {\Theta}_{3,3}(\vec{k}_1,\vec{k}_2)
+{\Theta}_{3,1}(\vec{k}_1,\vec{k}_2)
\bigg]
 \ \ + {\rm cyclic\ perms}.
\eea
Finally the dimensionless amplitude ${\hat {\cal A}}_{IV}$ is given by
\begin{eqnarray}
{\hat {\cal A}}_{IV}&=&  \sqrt{k_1 k_2} k_3  \int d^3 p \sqrt{p q}  \Big[ \frac{1}{k_3 ^2}+\frac{1}{p^2}+\frac{1}{q ^2} \Big] \Bigg \{ h(p,q)  \Big [{\Gamma}_1(\vec{k}_1,\vec{k}_2,\vec{p}) + {\Gamma}_3 (\vec{k}_1,\vec{k}_2,\vec{p}) \Big] \nonumber\\
&&- h(p,-q) \Big[ {\Gamma}_2 (\vec{q} ,\vec{k}_1, \vec{k}_2) + {\Gamma}_2  (-\vec{p},-\vec{k}_1,-\vec{k}_2)^* \Big] -  h(-p,q) \Big[ {\Gamma}_2 (\vec{p} ,\vec{k}_1, \vec{k}_2) + {\Gamma}_2  (-\vec{q},-\vec{k}_1,-\vec{k}_2)^* \Big ] \nonumber\\
&& + h(-p,-q) \Big[ {\Gamma}_3 (\vec{q} ,\vec{p}, \vec{k}_1) + {\Gamma}_1 (-\vec{p},-\vec{k}_1,-\vec{k}_2)^* \Big ] -  h(p,q) \Big[ {\Gamma}_2 (\vec{k}_1 ,\vec{k}_2, \vec{p}) + {\Gamma}_2 (\vec{k}_2,\vec{k}_1,\vec{p}) \Big ] \nonumber\\
&&-  h(p,-q) \Big[ {\Gamma}_3 (\vec{q} ,\vec{k}_1, \vec{k}_2) + {\Gamma}_3 (\vec{q},\vec{k}_2,\vec{k}_1) \Big ] + h(-p,q) \Big[ {\Gamma}_3 (\vec{p} ,\vec{k}_1, \vec{k}_2) + {\Gamma}_3 (\vec{p},\vec{k}_2,\vec{k}_1) \Big ] \nonumber\\
&&-  h(-p,-q) \Big[ {\Gamma}_2 (-\vec{k}_1 ,-\vec{k}_2, -\vec{p})^* +
  {\Gamma}_2 (-\vec{k}_2,-\vec{k}_1,-\vec{p})^* \Big ] \Bigg \}
\nonumber\\
&&+ \text{cyclic perms},
\end{eqnarray}
where as before ${\vec k}_3 = - {\vec k}_1 - {\vec k}_2$ and ${\vec q} = {\vec k}_3 - {\vec p}$.
Also we have defined the function $h(a,b)$ by
\bea
h(a,b) &=& \int_0^{\tau_0} d\tau \left[ e^{-i \tau(a + b + k_3)} - e^{-i \tau(a + b - k_3)} \right] \nonumber \\
&=&
\frac{i}{a + b - k_3} \left[ 1 - e^{-i \tau_0 (a + b - k_3)} \right] -
  \frac{i}{a + b + k_3} \left[ 1 - e^{-i \tau_0 (a + b + k_3)} \right].
\label{6.20}
\eea

\section{Order of magnitude estimates and discussion}

We now discuss the implications of our results for the bispectrum by
making some order of magnitude estimates.  The results depends on a
number of parameters and quantities:
\begin{itemize}

\item The ratio of the Hubble and Planck scales, $H/M_p$.

\item The slow roll parameters $\epsilon$ and $\eta$.  For our order
  of magnitude estimates we will assume that $\epsilon \sim \eta$.

\item In the isotropic case, the dependence on the wave vectors ${\vec
    k}_1$, ${\vec k}_2$ and ${\vec k}_3$ can be parameterized in terms
  of a dependence on the overall scale
$$
k \equiv (k_1 k_2 k_3)^{1/3}
$$
and the dependence on the two parameters $x_2 = k_2/k_1$ and $x_3 = k_3/k_1$ which
characterizes the ``shape'' of the bispectrum.

\item Some
conventional terminology for the various limits in the
  space of shape parameters $x_2$ and
  $x_3$ is as follows.
Conventionally one orders the
  momenta so that $x_3 < x_2 < 1$, and
the triangle inequality gives $x_2 + x_3 > 1$.  These inequalities
restrict the allowed configurations to a triangle in the $(x_2,x_3)$
plane, see, for example, Fig.\ 30 of Ref.\ \cite{baumann}.
The three corners of the triangle are the squeezed limit $x_3 \to 0$,
$x_2 \to 1$, the equilateral limit $x_3 \to 1$, $x_2 \to 1$, and the
folded limit $x_3 \to 1/2$, $x_2 \to 1/2$.  Two of the edges of the
triangle are the elongated limit $x_2 + x_3 \to 1$ with $0 \le x_3 \le
1/2$, and the isosceles limit $x_2 - x_3 \to 0$ with $1/2 \le x_3 \le
1$.


\item Suppose that ${\cal B}(k,x_2,x_3)/{\cal P}_{\cal R}(k)^2$ is
  proportional to a known function ${\cal S}(k,x_2,x_3)$, which is
  called as shape function.  Then we can define a dimensionless
  parameter $f_{\rm NL}$ for that shape function, a measure of the
  amplitude of the bispectrum.  The conventional definition is
  \cite{baumann}
\be
{\cal B}(k,x_2,x_3) = \frac{18}{5} f_{\rm NL
} {\cal P}_{\cal R}(k)^2
\frac{{\cal S}(k,x_2,x_3)}{{\cal S}(k,1,1)}.
\label{fNLdef}
\ee

\end{itemize}

\subsection{Generic initial states}
\label{sec:genericstates}

Let us start off by neglecting the shape dependence, assuming that
$k_1 \sim k_2 \sim k_3 \sim k$.
In this case the vacuum bispectrum (\ref{vac})
scales as ${\cal B}_{0,{\rm dr}} \sim H^4/(M_p^4 \epsilon k^6)$.
Suppose now that the occupation number of the initial state is of
order unity.  Suppose also that the initial state is ``generic'' in
the sense that all the dimensionless amplitudes ${\cal A}$ defined in
Sec.\ \ref{sec:bispectrum} are of order unity, so that, for example,
the three point function is not suppressed compared to the two point
function, etc.  If we are agnostic about the origin of the
non-Bunch-Davies initial state, then this genericity assumption is
well motivated.
However, specific scenarios for generating initial states
can give rise to suppression of the three point function with respect
to the two and four point functions, and violate our genericity
assumption.  For example, this occurs in the second class of secnarios
discussed in the introduction, where the non Bunch-Davies initial
states arise as a result of new physics at high energies which has
been integrated out.
See, for example, Ref.\ \cite{Jackson:2012qp}.
This important class of scenarios yields non-generic initial states.
Nevertheless, for the remainder of this subsection we will restrict
attention to generic initial states.

With the assumption that all the dimensionless amplitudes ${\cal A}$
are of order unity, we find that
the results (\ref{calBII}) and (\ref{calBIV}) for the contributions from
the two and four point functions are of the same order as the vacuum
contribution:
\be
{\cal B}_{II} \sim {\cal B}_{IV} \sim \frac{H^4}{M_p^4 \epsilon k^6}
\sim {\cal B}_{0,{\rm dr}}.
\ee
These contributions are therefore small and hard to detect.  By
contrast,
the contribution (\ref{calBIII}) from the three point function scales as
\be
{\cal B}_{III} \sim \frac{H^3}{M_p^3 \epsilon^{3/2} k^6}  \sim
\frac{M_p}{H \sqrt{\epsilon}} \  {\cal B}_{0,{\rm dr}}.
\label{enhanced00a}
\ee
We now eliminate $M_p/H$ in favor of $\epsilon$ and
the power spectrum $\Delta_{\cal R}$ using Eq.\ (\ref{4.3}).  This
gives
\be
{\cal B}_{III} \sim \frac{1}{\epsilon \sqrt{ 8 \pi^2 \Delta_{\cal
      R}^2}} {\cal B}_{0,{\rm dr}} \sim (3 \times 10^5) \ {\cal
  B}_{0,{\rm dr}},
\label{enhanced00}
\ee
where we have used the estimate $\epsilon \sim 0.01$ and the measured
value $\Delta_{\cal R} \sim 3 \times 10^{-5}$.

Thus, the dominant non-Gaussianity for  initial states with
$N_{\rm occ} \sim {\cal A}_{III} \sim1 $ is due to the three point function of the initial
state. This is easy to understand: a nonzero bispectrum is obtained from an
initial three point function from just the linear evolution, without requiring
any nonlinearities.  The contributions to the bispectrum from the
initial two and four point functions, on the other hand, are
suppressed since they require the
nonlinearities in the dynamics.
The enhancement factor of $M_p/(H \sqrt{\epsilon})$ in Eq.\ (\ref{enhanced00a}) was
previously obtained in a special case by Agarwal et al. \cite{agarwal}.
Note that this contribution to the bispectrum vanishes identically for
Gaussian initial
states, as studied in many previous investigations
\cite{porrati1,porrati2,holt,meer,meer1,ganc,shandera,kundu,das}.

The result (\ref{enhanced00}) could yield a large bispectrum,
detectable if ${\cal A}_{III} \gtrsim 10^{-3}$.
Assuming that ${\cal A}_{III} \sim N_{\rm occ}$,
the bispectrum could be detectable even
for initial occupation numbers small compared to unity, and $N_{\rm
  occ} \gtrsim 1$ is compatible with the backreaction constraint
(\ref{upper1}) for suitable values of the spectral index $n$ and
bandwidth $\chi$ \footnote{On the other hand $N_{\rm occ} \gtrsim 1$ would give order-unity
corrections to the power spectrum (\ref{ps0}).  This would disagree
with observations
unless the corrected spectrum is nearly scale invariant, which would
be a fine tuning.}.

It is not possible to give a generic prediction for the shape
dependence of this dominant piece ${\cal B}_{III}$ of the bispectrum,
since the shape dependence is just inherited from that of the initial
state.  Different scenarios for the origin of the non-vacuum initial
state will yield different shape dependences.

\subsection{Initial states with vanishing three point function}

We now specialize to initial states for which the three point function
is small or vanishing, for example Gaussian states
\cite{porrati1,porrati2,holt,meer,meer1,ganc,shandera,kundu,das}
or statistical mixtures of mode occupation number eigenstates
\cite{Parker}.  For such states ${\cal B}_{III}$ vanishes,
and we argued above that when $k_1 \sim k_2 \sim k_3$ and $N_{\rm occ}\sim 1$ we have
\be
{\cal B}_{II} \sim {\cal B}_{IV} \sim {\cal B}_{0,{\rm dr}},
\ee
so that the non-Gaussianity is small in this regime.  Nevertheless
it is still possible that $\mathcal{B}_{II}$ and/or $\mathcal{B}_{IV}$ can be large in
various limits where $k_1 \sim k_2 \sim k_3$ is violated or $N_{\rm occ} \gg 1 $.
It is not possible to estimate how ${\cal B}_{IV}$ scales in such
limits, since it depends on unknown properties of the connected part
of the initial four point function.  Therefore in the remainder of
this section we will restrict attention to states for which the bispectrum is dominated by ${\cal B}_{II}$.

\subsubsection{Large occupation number regime}

Let us first consider the possibilities for an enhanced bispectrum in
the regime $N_{\rm occ} \gg 1$,
assuming for simplicity that
$k_1 \sim k_2 \sim k_3 \sim k$.  In this regime
the vacuum bispectrum (\ref{vac})
scales as ${\cal B}_{0,{\rm dr}} \sim H^4/(M_p^4 \epsilon k^6)$,
and from Eq.\ (\ref{calBII}) we obtain
\be
{\cal B}_{II}/{\cal B}_{0,{\rm dr}} \sim \left[ N_{\rm occ}(k) \ \ \ \  \& \ \ \ \  N_{\rm occ}^2 (k) \right].
\label{7.1}
\ee
Here the notation inside the square brackets means that there are two
types of term that arise, terms
proportional to $N_{\rm occ}(k_1) \sim {\cal
  F}_1(k_1) \sim {\cal F}_2(k_1) \sim N_{\rm occ}(k_2) \sim N_{\rm
  occ}(k)$, and terms proportional to the products
$N_{\rm occ}(k_1) N_{\rm occ}(k_2) \sim N_{\rm occ}(k)^2$.

We see from Eq.\ (\ref{7.1}) that a bispectrum much larger than the Bunch-Davies bispectrum
requires large occupation numbers, for initial states with vanishing
three point function.  Occupation numbers that exceed unity are
compatible with the backreaction constraint.
However, as discussed in Sec.\
\ref{sec:back} above,
 large occupation numbers are only possible over
very narrow bandwidths, which is a kind of fine tuning.  In addition
occupation numbers in excess of unity over a range of wavenumbers
is only possible if the range of wavenumbers does not exceed about
one and half orders of magnitude, cf.\ Eq.\ (\ref{bandwdithbound1}) above.
Finally, as noted above, occupation numbers
$\gtrsim 1$ would give order-unity
corrections to the power spectrum (\ref{ps0}), which would disagree
with observations unless the corrected spectrum conspires to be nearly scale invariant. Taken together, these constraints imply that it is difficult for the
bispectrum to be significantly larger than the vacuum bispectrum in
this regime.


\subsubsection{Limiting shape regimes}
\label{sec:lsr}


Consider now squeezed triangle configurations, $k_1 \sim k_2 \gg
k_3$.
The vacuum bispectrum in this regime is, from Eq.\ (\ref{vac}),
\be
{\cal B}_{0,{\rm dr}} \sim \frac{H^4}{M_p^4 \epsilon k_1^3 k_3^3}.
\ee
The limiting behavior of ${\cal B}_{II}$ in this limit depends on
the direction in which the limit is approached, which can be along the
edge of the triangle corresponding to elongated configurations, or
from the interior of the triangle.  If we assume a direction from the interior,
the dominant piece of ${\cal B}_{II}$ is given by the
amplitude ${\hat {\cal A}}_{II}$.
We find from Eq.\ (\ref{hatcalAII}) that
\be
{\cal B}_{II} \sim \frac{H^4}{M_p^4 \epsilon k_1^3 k_3^3} \left\{ \frac{k_1}{k_3}\left[
  N_{\rm occ}(k_1) \ \ \ \&  \ \ \ N_{\rm occ}(k_1)^2  \ \ \
\& \ \ \ N_{\rm occ}(k_1) N_{\rm occ}(k_3) \right], \ \ \ N_{\rm occ} (k_3)  \right\}.
\label{ans1}
\ee
Here the notation inside the square brackets means that there are three
types of term that arise which scale $\propto k_3^{-4}$, terms
proportional to $N_{\rm occ}(k_1) \sim N_{\rm occ}(k_2) \sim {\cal
  F}_1(k_1) \sim {\cal F}_2(k_1)$, terms proportional to the products
$N_{\rm occ}(k_1)
N_{\rm occ}(k_2) \sim N_{\rm occ}(k_1)^2$, and terms proportional to
the product $N_{\rm occ}(k_1) N_{\rm occ}(k_3)$. Also there  are terms proportional
to $\mathcal{F}_1 (k_3) \sim \mathcal{F}_2 (k_3) \sim N_{\rm occ}(k_3)$ which scale $\propto k_3^{-3}$.

If we now assume that all the occupation numbers are of order unity,
following Agullo and Parker \cite{Parker},
then we obtain
\be
{\cal B} / {\cal B}_{0,{\rm dr}} \sim k_1/k_3.
\label{large}
\ee
Since $k_1/k_3$ can be as large as $\sim 100$ for modes probed in the
CMB, and even larger for modes probed by large scale structure, the
estimate (\ref{large}) is a significant enhancement of non-Gaussianity
over the standard vacuum result, as previously argued in a special
case by Agullo and Parker \cite{Parker}, and by Ganc \cite{ganc} for
the the class of squeezed vacuum states.

However the assumption that all mode occupation numbers are of order
unity is quite restrictive and can conflict with the backreaction
constraints discussed in Sec.\ \ref{sec:back}.
We now investigate how large the
enhancement factor (\ref{large}) can be assuming the power law model (\ref{pl}) is
valid over all scales from $k_3$ to $k_1\sim k_2$.
We will find that
the backreaction constraints imply that the enhancement factor
cannot exceed $\sim 200$.

The backreaction constraints will be weakest when
$k_{\rm min} \sim k_3$ and $k_{\rm max} \sim k_1$, so we will assume
this in what follows.
Consider first the first term in square brackets in Eq.\ (\ref{ans1}),
$N_{\rm occ}(k_1)$, which is constrained by Eq.\ (\ref{N1bound}).
Combining Eqs.\ (\ref{large}) and (\ref{N1bound}) gives that this
contribution to the bispectrum is bounded above by
${\cal B}_{II}/{\cal B}_{0,{\rm dr}} \lesssim 10^5 n / (\chi^3
(1-\chi^{-n}))$, which is of order unity or smaller for $|n| \le 10$
and $\chi = k_{\rm max}/k_{\rm min} \sim k_1/k_3 \gtrsim 10^2$.
A similar argument applies to the second term in the square brackets
in Eq.\ (\ref{large}).  The largest term is the third term, which
from Eqs.\ (\ref{upper1}), (\ref{N1bound}) and (\ref{large}) is
bounded above by
\be
\frac{{\cal B}_{II}}{{\cal B}_{0,{\rm dr}}} \sim \chi N_0 N_1 \lesssim
\frac{10^{10} \chi^{n-3}}{f_n(\chi)^2} \sim \frac{10^{10} \chi^{n-3}
  n^2}{(\chi^n-1)^2} \lesssim \frac{10^{10}}{ \chi^3 (\ln \chi)^2},
\label{upper3}
\ee
where we have used the fact that the maximum of the function of $n$ is
achieved at $n=0$.
The upper bound (\ref{upper3}) can be quite large for $\chi \sim 1$,
but that regime corresponds to measuring the bispectrum over a narrow
range of scales, and choosing the initial state to populate only those
scales, which would be a considerable fine tuning.
The the upper bound (\ref{upper3}) is smaller than $\chi$,
invalidating the estimate (\ref{large}), when $\chi
\gtrsim 200$.

We now turn to a discussion of the
the edge of the triangle corresponding to
elongated configurations, i.e.\ the line $k_1=k_2 +
k_3$.
There is an enhancement of the bispectrum for these
configurations.
For the special case of squeezed states
(cf.\ Sec.\ \ref{sec:multi} below), this enhancement was previously pointed out by Chen {\it et. al.} \cite{Chen:2006nt}
and Holman and Tolley \cite{holt},
and studied in more detail by Agullo and Shandera \cite{shandera}.
The enhancement arises
from the four terms in Eq.\ (\ref{hatcalAII}) in which one of the arguments
of the function $r(a,b)$ is $\pm {\tilde k}_1$.
Normally the phases $k \tau_0$ in the formula (\ref{rdef}) for
$r(a,b)$ have absolute values which are large compared to unity, since
we have assumed that all the modes are inside the horizon at the
initial time, $k_i |\tau_0| \gg 1$. Therefore the function $r$ scales
as $r \sim 1/k$.  However, for elongated configurations, ${\tilde k}_1
\to 0$ from Eq.\ (\ref{tildekdef}), and so for example the function $r({\tilde k}_1, -{\tilde
  k}_2)$ scales as $\sim |\tau_0|$, enhanced by a factor $k |\tau_0|
\gg 1$.
We find from Eqs.\ (\ref{hatcalAII}) and (\ref{rdef})
that in the elongated limit\footnote{The same estimate also applies to
  the squeezed limit approached along the edge of the triangle, i.e.\
  via elongated configurations.}
\be
\frac{\mathcal{B}_{II}}{\mathcal{B}_{\rm 0,dr}} \sim |\tau_0| k_1
W[k_1 |\tau_0| (x_2 + x_3 -1)]
\big [ N_{\rm occ} \ \ \& \ \ N_{\rm occ}^2 \big ]
,
\label{elongated}
\ee
where $W(x) = (1 - e^{-i x})/(ix)$.  The factor $W[k_1 |\tau_0| (x_2 +
x_3 -1)]$ goes to unity along the edge $k_2 + k_3 = k_1$ of
the triangle.

How large can the enhancement factor (\ref{elongated}) be?
The factors of occupation number are limited by the backreaction
constraint, just as discussed above for squeezed configurations.
The function $W$ is of order unity or smaller.
The factor $k_1 |\tau_0|$ can apparently be arbitrarily large,
since the initial conformal time $\tau_0$ can be chosen to be as early
as desired.  However, this large $\tau_0$ divergence is
only apparent, it is an
artifact\footnote{This aspect was missed in the analysis of Chen {\it et. al.} \cite{Chen:2006nt} and Holman and
Tolley \protect{\cite{holt}}.}  of how we
have chosen to parameterize the initial state, since it diverges as
$\tau_0 \to -\infty$.  If we specify the initial state at some time
$\tau_0$, then changing the initial time to some earlier time $\tau_1$
will have no affect on the bispectrum if we choose the initial state
at $\tau_1$ to be that obtained from the initial state at $\tau_0$ by
evolving
backwards in time from $\tau_0$ to $\tau_1$.  Our simple power law
model (\ref{est}) and (\ref{pl}) for the initial state is not invariant
under time evolution when interactions are included, which is the
reason for the $\tau_0$ dependence of the bispectrum (\ref{finalresult}).
To avoid this spurious time dependence, we will choose the value of $\tau_0$ to be
the latest possible time for which all the modes under consideration
are inside the horizon.  This choice gives $|\tau_0| \sim 1/k_{\rm
  min}$, so $|\tau_0| k_1 \sim |\tau_0| k_{\rm max} \sim k_{\rm
  max}/k_{\rm min} \sim \chi$.
With this choice the enhancement factor (\ref{elongated}) for
elongated configurations is of the same order as the enhancement
factor (\ref{upper3}) for squeezed configurations
approached from the interior
of the triangle.

\subsubsection{Observational constraints}

We now discuss observational constraints on the class of states under
discussion in this section, where the bispectrum is dominated by the initial two
point function.
As discussed above it is not possible to make detailed predictions for
the cases where the bispectrum is dominated by the initial three or
four point functions.

For simplicity we make the following assumptions and specializations.
We assume that the functions ${\cal F}_1$ and ${\cal F}_2$ are real
and satisfy ${\cal F}_1 = {\cal F}_2 = N_{\rm occ}$.  For the
occupation number $N_{\rm occ}$ we use the model (\ref{pl}) specialized
to $n=4$.
Then the result (\ref{calBII}) for the bispectrum simplifies to
\begin{eqnarray}
\mathcal{B}_{II} \approx  \frac{18}{5}\ \mathcal{P}^2 (k) \bigg(\frac{20}{3} N_0 ^2 \epsilon\bigg) \frac{1}{3 k^3}  \Bigg[  ( k_1
^2 k_2 ^2 + k_3 ^2 k_1 ^2 + k_3 ^2 k_2 ^2) \sum_{i = 1}^{3} \frac{1- \cos{\tilde{k_i}\tau_0}}{\tilde{k_i}} \Bigg],
\label{6.5}
\end{eqnarray}
where we have omitted terms proportional to $(1-\cos{k_t \tau_0})/k_t$ by virtue of being subdominant to other terms 
 in triangular regimes of interest.
Also given our specialization $\mathcal{F}_1 = \mathcal{F}_2$, the power spectrum above is just the  Bunch-Davies  power spectrum introduced in Sec. \ref{sec:power}.
From Eq. \eqref{6.5} and the definition of $f_{\text{NL}}$ provided in Eq. \eqref{fNLdef} we have
\begin{equation}
f_{\text{NL}} = \frac{20}{3} N_0 ^2 \epsilon.
\end{equation}
The square bracket on the right of Eq. \eqref{6.5}  includes the shape functions
\begin{eqnarray}
&&\mathcal{S}^{1} _{\text{NBD}} \equiv \frac{1}{2} k_1 ^2 (k_2^2+ k_3 ^2) \  \frac{1- \cos{\tilde{k_1}\tau_0}}{\tilde{k_1}} + (\text{cyclic  perm.}) 
\label{anz1}
\end{eqnarray}
and
\begin{eqnarray}
&& \mathcal{S}^{2} _{\text{NBD}} \equiv (k_2 ^2 k_3 ^2) \  \frac{1- \cos{\tilde{k_1}\tau_0}}{\tilde{k_1}} + (\text{cyclic  perm.})  
\label{anz2}
\end{eqnarray}
which were first introduced as shape function ansatzs by Parker and
Agullo\cite{Parker}.  Here ``NBD'' stands for non Bunch-Davies. They
have been analyzed by the Planck collaboration \cite{planck23}, who
obtained
upper bounds on the corresponding $f_{\rm NL}$ parameters which they define as 
\begin{eqnarray}
&&\mathcal{B}_{\text{NBD}} =  \frac{2}{k^3} \ \mathcal{P}^{2} (k) f^{\text{NBDi}} _{\text{NL}} \ \mathcal{S}_{\text{NBD}} ^{i}.
\end{eqnarray}
The parameter $f_{\rm NL}^{\rm NBD1}$ has a  shape that peaks in squeezed
configurations, while the parameter $f_{\rm NL}^{\rm NBD2}$ has a 
shape that peaks in folded configurations.
Comparing our bispectrum in Eq. \eqref{6.5} to the Bunch-Davies bispectrum ansatz 
in Eqs. \eqref{anz1} and \eqref{anz2}  we find

\be
f_{\rm NL}^{\rm NBD1} = \frac{6}{5} f_{\rm NL}, \hspace{1cm} f_{\rm NL}^{\rm NBD2} = \frac{27}{20} f_{\rm NL}.
\ee

The Planck collaboration's upper bounds are given in Table 11 of Ref.\ \cite{planck23} as
\be
f_{\text{NL}} ^{\text{NBD1}} =  31 \pm 19, \ \ \ \ f_{\text{NL}} ^{\text{NBD2}} = 0.8 \pm 0.2.
\label{bounds}
\ee
The corresponding constraint on the initial occupation number $N_0$ obtained from
$f_{\rm NL} ^{\rm NBD2}$ is more stringent. Specifically, for $\epsilon
\sim 0.01$, an occupation number $N_0 \approx 3$ is obtained which is consistent with
the bounds (\ref{bounds}).  This value
is however large compared to limits from backreaction or constancy of power
spectrum considerations. In the same way, the Planck data does not
strongly constrain the bandwidth $\chi$.

\subsection{Example: multimode squeezed initial states}
\label{sec:multi}

An example of a class of initial states to which our analysis applies
are multimode squeezed states or generalized vacuum states.  These are defined by considering
mode operators ${\hat B}_{\vec{k}}$ which are related to the operators
${\hat A}_{\vec{k}}$ of the mode expansion (\ref{2.9}) by a linear
transformation of the form
\be \begin{pmatrix}
\hat{B}_{\vec{k}} \\
\hat{B}^{\dagger} _{-\vec{k }}\\
\end{pmatrix}   = \begin{pmatrix} \alpha^* (k)    &  -\beta^* (k)\\
-\beta(k) & \alpha(k) \\   \end{pmatrix} \begin{pmatrix} \hat{A} _{\vec{k}} \\  \hat{A}^{\dagger}_{-\vec{k}}\\ \end{pmatrix} ,
\label{squeezed}
\ee
where $ |\alpha(k)|^2 - |\beta(k)|^2 = 1$.  The squeezed state at the
initial time $\tau_0$ is then the state that is annihilated by the
operators ${\hat B}_{\vec{k}}$.  This is the class of states that
would be produced by starting at an earlier time with a Bunch-Davies
vacuum, and then evolving through a homogeneous isotropic expansion
phase with a single scalar field.
Non-Gaussianities produced by this class of states have been studied
in Refs.\ \cite{porrati1,porrati2,holt,meer,meer1,ganc,shandera,Ashoorioon:2013eia,Aravind:2013lra}.

For this class of states, we can compute the free functions ${\cal F}_i$, ${\cal H}_i$ and
$\Gamma_i$ that parameterize the initial state
by inserting the definition (\ref{squeezed})
into Eqs.\
(\ref{3.1and2}) -- (\ref{decom}).  This gives
\be
{\cal H}_i = \Gamma_i = 0,
\ee
and
\be
{\cal F}_2({\vec k}) = |\beta(k)|^2, \ \ \ \ {\cal F}_1({\vec k}) = \alpha(k) \beta(k)^*.
\label{7.13}
\ee
Inserting these functions into our expression (\ref{calBII}) for the
bispectrum reproduces the result obtained previously by Ganc
\cite{ganc}, and earlier in a certain limit by Chen {\it et al.}
\cite{Chen:2006nt}.
Similarly, inserting these functions into our
expression (\ref{ps0}) for the leading order corrections to the power
spectrum reproduces previous estimates of the effects of
trans-Planckian physics by Easther {\it et. al.} \cite{egks02}, for
their choice of Bogolubov coefficients
\be
\alpha = 1+ x \frac{H}{\Lambda} + \mathcal{O}\Big( \frac{H^2}{\Lambda ^2} \Big), \ \ \ \
\beta = y \frac{H}{\Lambda} + \mathcal{O}\Big( \frac{H^2}{\Lambda ^2} \Big),
\label{7.16}
\ee
where $\Lambda$ is a cutoff.

\section{Acknowledgments}

We thank Liam McAllister for helpful discussions. This research was supported in part by
NSF grants PHY-1068541 and PHY-0968820 and by NASA grant
NNX11AI95G.

\appendix

\section{Free functions for the dressed vacuum state}
\label{app:dressedvacuum}

In this appendix we compute the free functions ${\cal F}_i$, ${\cal
  H}_i$ and $\Gamma_i$ that characterize the initial state, for the
ground state of the interacting theory, the dressed vacuum.

We are interested in evaluating the initial two point, three point and four
point functions.  The general $n$-point function is
\be
\langle \Omega | \hat{A}^{\alpha}_{\vec{k}_1}
... \hat{A}^{\delta}_{\vec{k}_n} | \Omega \rangle,
\label{npoint}
\ee
where $| \Omega \rangle$ is the dressed vacuum state.
Here the indices $\alpha$, $\beta$ etc. can be $0$ or $1$; the
notation is that ${\hat A}^\alpha$ means ${\hat A}$ for $\alpha=0$ and
${\hat A}^\dagger$ for $\alpha=1$.  Note that the $n$-point function
(\ref{npoint}) is time dependent.  In the Schr\"odinger picture the
state $|\Omega\rangle$ is time independent while the operators ${\hat
  A}_{\vec k}$ are time dependent, while in the interaction picture
the converse is true.  We will evaluate this quantity at the initial
conformal time $\tau_0$.

As usual we can evolve expectation values in the dressed vacuum by
evolving starting with the free vacuum at early times, with the early
time displaced slightly into the complex plane \cite{Peskin:257493,maldacena}:
\be
\langle \Omega | \hat{A}^{\alpha}_{\vec{k}_1}
... \hat{A}^{\delta}_{\vec{k}_n} | \Omega \rangle =  \lim_{\tau
  \rightarrow -\infty (1- i \epsilon)}\langle 0 |
{\hat U}(\tau_0,\tau)^\dagger
\hat{A}^{\alpha}_{\vec{k}_1} ... \hat{A}^{\delta}_{\vec{k}_n} {\hat U}(\tau_0,\tau) | 0 \rangle.
\label{dcomplex}
\ee
Here ${\hat U}(\tau_0,\tau)$ is the unitary evolution operator that
maps interaction picture states at conformal time $\tau$ to conformal
time $\tau_0$, given by
\be
{\hat U}(\tau_0,\tau) = T \exp\left\{-i \int_\tau^{\tau_0}
d{\tau'} {\hat H}_{\rm int}^I(\tau')\right\}.
\ee
Also in Eq.\ (\ref{dcomplex}) the operators ${\hat A}_{\vec k}$
are interaction picture operators; we drop the I denoting
interaction picture from now on.
Expanding to leading order in the interaction gives
\be
\langle \Omega | \hat{A}^{\alpha}_{\vec{k}_1}
... \hat{A}^{\delta}_{\vec{k}_n} | \Omega \rangle =
\langle 0 | \hat{A}^{\alpha}_{\vec{k}_1}
... \hat{A}^{\delta}_{\vec{k}_n} | 0 \rangle +
i \int_{-\infty (1-
  i \epsilon)} ^{\tau_0} d \tau   \left< 0 \left| \left[ {\hat
        H}_{\text{int}}(\tau),  \hat{A}^{\alpha}_{\vec{k}_1}
      ... \hat{A}^{\delta}_{\vec{k}_n}\right] \right|0 \right>.
\label{ggg}
\ee

Next, we use the explicit formula (\ref{inter}) for the interaction
Hamiltonian.  Since this interaction Hamiltonian is trilinear in the
fields,
and since the vacuum expectation value of the product of an odd number of
creation or annihilation operators vanishes, it follows from Eq.\ (\ref{ggg})
that all of
the free functions (\ref{3.1and2}) and (\ref{calGdef}) vanish.
The only possible non-vanishing free functions are ${\cal H}_1$ and
${\cal H}_2$, defined by Eq. (\ref{calHdef}).  We now
evaluate $\mathcal{H}_1 ^{0,\text{dr}}$ to leading order:
\bea
\left< \Omega | \hat{A}_{\vec{k}_1} \hat{A}_{\vec{k}_2}
  \hat{A}_{\vec{k}_3} | \Omega
\right>  &=&
(2 \pi)^3 \delta(\vec{k}_1+\vec{k}_2+\vec{k}_3) \mathcal{H}_1
^{0,\text{dr}} (\vec{k}_1,\vec{k}_2) \nonumber \\
&=&  i \int_{-\infty (1- i
  \epsilon)} ^{\tau_0} d \tau   \left< 0 | \left[ {\hat H}_{\text{int}},  \hat{A}_{\vec{k}_1}  \hat{A}_{\vec{k}_2} \hat{A}_{\vec{k}_3} \right] |0 \right>  \nonumber\\
 &=& - \frac{\sqrt{\epsilon} H}{2 M_p}
\int d\tau
\int \frac{d^3k'}{(2 \pi)^3} \int \frac{d^3 p}{(2 \pi)^3} \frac{\sqrt{k' |\vec{p}-\vec{k}'| p}}{p^2}
\left< 0 \big| \left[ (\hat{A}_{\vec{k}'} e^{-i \tau k'} -
    \hat{A}^{\dagger}_{-\vec{k}'} e^{i \tau k'}) \right. \right.
\nonumber\\
&&
\left. \left.
(\hat{A}_{\vec{p}-\vec{k}'} e^{-i \tau |\vec{p}-\vec{k}'| } - \hat{A}^{\dagger}_{\vec{p}-\vec{k}'} e^{i \tau  |\vec{p}-\vec{k}'| })(\hat{A}_{-\vec{p}} e^{-i \tau p} - \hat{A}^{\dagger}_{\vec{p}} e^{i \tau p}), \hat{A}_{\vec{k}_1}  \hat{A}_{\vec{k}_2} \hat{A}_{\vec{k}_3} \right] \big| 0 \right>
\nonumber\\
&=& -\frac{\sqrt{\epsilon} H}{2 M_p} \int_{-\infty (1- i \epsilon)}
^{\tau_0} d \tau \int \frac{d^3k'}{(2 \pi)^3} \int \frac{d^3 p}{(2
  \pi)^3} \frac{\sqrt{k' |\vec{p}-\vec{k}'| p}}{p^2}
\nonumber \\ &&
\times \left< 0 \big|  \hat{A}_{\vec{k}_1}  \hat{A}_{\vec{k}_2}
\hat{A}_{\vec{k}_3}  \hat{A}^{\dagger}_{-\vec{k}'}
\hat{A}^{\dagger}_{\vec{p}-\vec{k}'} \hat{A}^{\dagger}_{\vec{p}} \big|
0 \right>
 e^{i \tau( k' + |\vec{p}-\vec{k}'| + p)}.
\label{A.16}
\end{eqnarray}
Here we have used the interaction Hamiltonian (\ref{inter}), the mode
expansion (\ref{2.9a}) and the asymptotic form (\ref{2.13}) of the mode functions.
Simplifying Eq.\ (\ref{A.16}) gives
\be
 \mathcal{H}_1 ^{0,\text{dr}} (\vec{k}_1,\vec{k}_2) = i \frac{H}{M_p} \sqrt{\epsilon} \frac{(k_1 k_2 k_3)^{1/2}}{k_t} \sum_{i=1}^3 \frac{1}{k_i ^2 } e^{i \tau_0 k_t},
\label{dressed}
\ee
where $\vec{k}_3 = - \vec{k}_1-\vec{k}_2$ and $k_t = k_1 + k_2 + k_3$.
A similar calculation shows that ${\cal H}_2^{0,\text{dr}}$ vanishes.

We note that there is no contribution to these computations from the
field redefinition (\ref{5.2}), since the operators ${\hat A}_{\vec k}$
are defined in terms of the mode expansion (\ref{2.9a}) of the redefined field
operator ${\hat {\cal R}}_c$.

\section{Explicit calculation of the bispectrum}
\label{app:bispectrum}

We fix values of ${\vec k}_1$, ${\vec k}_2$ and ${\vec k}_3$, all of
whom we assume to be nonvanishing.  As discussed near Eq.\ (\ref{5.4}),
there are two contributions to bispectrum, one from the field
redefinition (\ref{5.2}) and one from the interaction Hamiltonian.
We now compute these two contributions in turn.

\subsection{Field redefinition contribution}

We substitute the field redefinition (\ref{5.2}) into the first term
on the right hand side of Eq.\ (\ref{5.4}).  We neglect the second
term in Eq.\ (\ref{5.4}) for the moment; that term will be treated in
Sec.\ \ref{sec:ccc} below.
The result is
\bea
\label{longresult}
\langle  \hat{\mathcal{R}}^{I} _{\vec{k}_1}\hat{ \mathcal{R}}^{I}_{\vec{k}_2} \hat{\mathcal{R}}^{I}_{\vec{k}_3}  \rangle
& &=\langle \hat{\mathcal{R}}^{I}_{c,\vec{k}_1}
 \hat{\mathcal{R}}^{I}_{c,\vec{k}_2}
 \hat{\mathcal{R}}^{I}_{c,\vec{k}_3}  \rangle
 \\
&&+ \left( \frac{1}{2} \frac{\ddot{\phi}}{\phi H} + \frac{1}{8
    M^2_{p}} \frac{\dot{\phi}^2 }{H^2}\right)  \left( \int
\frac{d^3 p}{(2 \pi)^3} \left<  \hat{\mathcal{R}}^{I}_{c,\vec{k}_1}
\hat{\mathcal{R}}^{I}_{c,\vec{k}_2}  \hat{\mathcal{R}}^{I}_{c,\vec{p}}
\hat{\mathcal{R}}^{I}_{c,\vec{k}_3- \vec{p}}  \right> + \text{cyclic
perms} \right) \nonumber \\
&&  +  \frac{\dot{\phi} ^2}{4 M_p^2 H^2}  \left(
\frac{1}{k^2 _3 } \int \frac{d^3 p}{(2 \pi)^3} \left<
\hat{\mathcal{R}}^{I}_{c,\vec{k}_1}
\hat{\mathcal{R}}^{I}_{c,\vec{k}_2} \hat{\mathcal{R}}^{I}_{c,\vec{p}}
\hat{\mathcal{R}}^{I}_{c,\vec{k}_3- \vec{p}} \right> (\vec{k}_3 -
\vec{p} )^2 + \text{cyclic perms} \right). \nonumber
\eea
Using the mode expansion (\ref{2.9a}), the asymptotic form
(\ref{2.13}) of the mode functions,
the decomposition (\ref{hatdef}) of the function ${\cal H}_1$,
and noting that one point
functions vanish because of our homogeneity assumption, we find
for the first term in Eq.\ (\ref{longresult})
\be \langle
\hat{\mathcal{R}}^I_{c,\vec{k}_1}\hat{\mathcal{R}}^I_{c,\vec{k}_2}
\hat{\mathcal{R}}^I_{c,\vec{k}_3}  \rangle = (2 \pi)^3
\delta(\vec{k}_1+\vec{k}_2+\vec{k}_3)\left[ \mathcal{B}_{III}
(\vec{k}_1,\vec{k}_2,\vec{k}_3) + \mathcal{B}_{III}
(\vec{k}_1,\vec{k}_2,\vec{k}_3)^{0,\text{dr}}\right].
\label{eq:B3}
\ee
Here
\bea
 \mathcal{B}_{III} (\vec{k}_1,\vec{k}_2,\vec{k}_3)  &=&  -\frac{i H^3
 }{8 M^3 _{p} \epsilon ^{\frac{3}{2}}} \frac{1}{(k_1 k_2
   k_3)^{\frac{3}{2}}} \left[\hat{\mathcal{H}}_1 (\vec{k}_1,\vec{k}_2)
   - \hat{\mathcal{H}}_1 (-\vec{k}_1,-\vec{k}_2)^*-\mathcal{H}_2
   (\vec{k}_1,\vec{k}_2)
\right.
\nonumber \\
&&
+\mathcal{H}_2 (-\vec{k}_1,-\vec{k}_2)^*
-\mathcal{H}_2 (\vec{k}_1,\vec{k}_3) +\mathcal{H}_2
(-\vec{k}_1,-\vec{k}_3)^*    -\mathcal{H}_2 (\vec{k}_2,\vec{k}_3)
\nonumber \\ &&
\left. +\mathcal{H}_2 (-\vec{k}_2,-\vec{k}_3)^*   \right]
\eea
and
\bea
\label{B.6}
\mathcal{B}_{III} (\vec{k}_1,\vec{k}_2,\vec{k}_3)^{0,\text{dr}} &=&
-\frac{i H^3 }{8 M^3 _{p} \epsilon ^{\frac{3}{2}}} \frac{1}{(k_1 k_2
  k_3)^{\frac{3}{2}}} \big [\mathcal{H}^{0,\text{dr}}_1
(\vec{k}_1,\vec{k}_2) - \mathcal{H}^{0,\text{dr}}_1
(-\vec{k}_1,-\vec{k}_2)^* \big ] \nonumber \\
&&
 = \frac{ H^4 }{4 M^4 _{p}}   \frac{\epsilon}{k_1 k_2 k_3} \sum_{i=0}^3 \frac{1}{k_i^2 k_t} \cos(\tau_0 k_t),
\eea
where we used the expression (\ref{dressed}) for
$\mathcal{H}_1^{0,{\rm dr}}$.

We can similarly evaluate the products of four field operators that
appear in Eq.\ (\ref{longresult}).  We find that
$\langle  \hat{\mathcal{R}}_{\vec{k}_1} \hat{\mathcal{R}}_{\vec{k}_2}
\hat{\mathcal{R}}_{\vec{p}} \hat{\mathcal{R}}_{\vec{k}_3- \vec{p}}
\rangle$ evaluates to
\bea
&&(2 \pi)^6 \delta(\vec{k}_1+\vec{k}_2+\vec{k}_3) \frac{H^4 }{16 M^4
  _{p}   \epsilon ^2 }  \frac{1}{(k_1 k_2 p | \vec{k}_3 -\vec{
    p}|)^{\frac{3}{2}}} \Bigg [\mathcal{G}_1
(\vec{k}_1,\vec{k}_2,\vec{p})- \mathcal{G}_2
(\vec{k}_3-\vec{p},\vec{k}_1,\vec{k}_2)-\mathcal{G}_2
(\vec{p},\vec{k}_1,\vec{k}_2)
\nonumber \\ &&
-\mathcal{G}_2
(\vec{k}_2,\vec{k}_1,\vec{p})-\mathcal{G}_2(-\vec{k}_1,-\vec{k}_2,-\vec{p})^*-\mathcal{G}_2(\vec{k}_1,\vec{k}_2,\vec{p})
-\mathcal{G}_2(-\vec{k}_2,-\vec{k}_1,-\vec{p})^*
-\mathcal{G}_2(-\vec{p},-\vec{k}_1,-\vec{k}_2)^*
\nonumber \\ &&
-\mathcal{G}_2(\vec{p}-\vec{k}_3,-\vec{k}_1,-\vec{k}_2)^*+\mathcal{G}_3
(\vec{p},\vec{k}_3-\vec{p},\vec{k}_1)
+\mathcal{G}_3(\vec{k}_2,\vec{k}_3-\vec{p},\vec{k}_1) +\mathcal{G}_3
(\vec{k}_2,\vec{p},\vec{k}_1)+\mathcal{G}_3(\vec{k}_1,\vec{k}_3-\vec{p},\vec{k}_2)
\nonumber \\ &&
+ \mathcal{G}_3(\vec{k}_1,\vec{p},\vec{k}_2)
+\mathcal{G}_3(\vec{k}_1,\vec{k}_2,\vec{p})
+\mathcal{G}_1(-\vec{k}_1,-\vec{k}_2,-\vec{p})^*
\nonumber \\ &&
  +  \delta(\vec{p}+\vec{k}_1) \Big \{ -2 \text{Re} \mathcal{F}_1(\vec{k}_1)  - 2 \text{Re} \mathcal{F}_1(\vec{k}_2)+ \mathcal{F}_2 (-\vec{k}_2) + \mathcal{F}_2(-\vec{k}_1) +\mathcal{F}_2(\vec{k}_2)+\mathcal{F}_2(\vec{k}_1)\Big \}
\nonumber \\ &&
+ \delta (\vec{p}+\vec{k}_2) \Big \{-2 \text{Re} \mathcal{F}_1(\vec{k}_1) - 2 \text{Re} \mathcal{F}_1(\vec{k}_2) + \mathcal{F}_2 (-\vec{k}_2) + \mathcal{F}_2(-\vec{k}_1) +\mathcal{F}_2(\vec{k}_2)+\mathcal{F}_2(\vec{k}_1) \Big \}
\nonumber \\ &&
 +  \delta(\vec{p}+\vec{k}_1) + \delta(\vec{p}+\vec{k}_2)  \Bigg].
\eea
We now substitute this expression together with the expression
(\ref{eq:B3}) back into Eq.\ (\ref{longresult}), and then into Eq.\
(\ref{5.4}),
and then simplify using the decompositions (\ref{decom}).
This yields the following pieces of our expression (\ref{finalresult}) for the
bispectrum: a part of the vacuum term ${\cal B}_{0,{\rm dr}}$, the
term ${\cal B}_{III}$, the pieces ${\cal A}_{II}$ and
${\tilde {\cal A}}_{II}$ of the term ${\cal B}_{II}$, and the pieces
${\cal A}_{IV}$ and ${\tilde {\cal A}}_{IV}$ of the term ${\cal
  B}_{IV}$.  It also generates the additional term (\ref{B.6})
which will be canceled by a term obtained in the next subsection.

\subsection{Commutator contribution}
\label{sec:ccc}

We now evaluate the second term in Eq.\ (\ref{5.4}).
For this term the effect of the field redefinition is
of subleading order and can be neglected.
Using the interaction Hamiltonian (\ref{inter}), the mode
expansion (\ref{2.9a}), and the asymptotic form (\ref{2.13}) of the
mode functions
we find that this term evaluates to
\bea
\label{secondterm}
&&
  - \int_{\tau_0}^0 d\tau' \frac{i H^4}{16 \epsilon M_p ^4 }
\Big(\frac{1}{k_1 k_2 k_3 } \Big)^{\frac{3}{2}} \int \frac{d^3 p}{(2
  \pi)^3}  \frac{d^3 k'}{(2 \pi)^3} \frac{\sqrt{p k'
    |\vec{p}-\vec{k}'|}}{p^2}
\left< \left[ \left(\hat{A}_{\vec{k}_1} - \hat{A}^{\dagger} _{-\vec{k}_1} \right)
    \left(\hat{A}_{\vec{k}_2} -\hat{ A}^{\dagger} _{-\vec{k}_2} \right) \right.\right.
\nonumber \\ &&
\times    \left(\hat{A}_{\vec{k}_3} - \hat{A}^{\dagger} _{-\vec{k}_3} \right) ,
 \left(\hat{A}_{\vec{k}'}  e ^{-i \tau' k'} -\hat{
     A}^{\dagger}_{-\vec{k}'}e^{i \tau' k'} \right)
\left(\hat{A}_{\vec{p}- \vec{k}'}  e ^{-i \tau'| \vec{p}-\vec{k}'| } -
  \hat{A}^{\dagger} _{\vec{k}'-\vec{ p }}e^{i \tau'|\vec{p}-\vec{k}'|}
\right)
\nonumber \\ &&
\left. \left. \times \left(\hat{A}_{-\vec{p}}  e ^{-i \tau' p} -
  \hat{A}^{\dagger}_{\vec{p}}e^{i \tau' p} \right) \right] \right>.
\eea
Using the definitions (\ref{3.1and2}) and (\ref{calGdef}) of the free functions
${\cal F}_i$ and ${\cal G}_i$, the expression inside the expectation
value brackets $\left< \ldots \right>$ in Eq.\ (\ref{secondterm}) evaluates to
\begin{eqnarray}
\label{B7}
&& (2 \pi)^9  \delta(\vec{p}-\vec{k}_3) \delta(\vec{k}_1+\vec{k}_2+\vec{p}) \Bigg [ (e^ { -i \tau'  (k' + | \vec{p}- \vec{k}'| + p)} -  e^ { -i \tau'  (k' + | \vec{p}-\vec{k}'| - p)} ) \big ( \mathcal{G}_1(\vec{k}_1,\vec{k}_2,\vec{k}') + \mathcal{G}_3(\vec{k}_2,\vec{k}_1,\vec{k}') \big) \nonumber\\
&&+   (e^ { -i \tau'  (k' - | \vec{p}-\vec{k}'| - p)} -  e^ { -i \tau'  (k' - | \vec{p}-\vec{k}'| + p)} )   \big ( \mathcal{G}_2 (\vec{p}-\vec{k}',\vec{k}_1,\vec{k}_2) + \mathcal{G}_2  (-\vec{k}',-\vec{k}_1,-\vec{k}_2)^* \big )  \nonumber\\
&&+ (e^ { -i \tau' (-k' + | \vec{p}- \vec{k}'| - p)} -  e^ { -i \tau'  (-k' + | \vec{p}-\vec{k}'| + p)} ) \big (\mathcal{G}_2 (\vec{k}',\vec{k}_1,\vec{k}_2)+\mathcal{G}_2  (\vec{k}'-\vec{p},-\vec{k}_1,-\vec{k}_2 )^* \big ) \nonumber\\
&&+ (  e^ { -i \tau'  (-k'  -| \vec{p}- \vec{k}'| + p)}- e^ { -i \tau'  (-k' - | \vec{p}-\vec{k}'| - p)}  ) \big( \mathcal{G}_3 (\vec{p}-\vec{k}',\vec{k}',\vec{k}_1) + \mathcal{G}_1  (-\vec{k}_1,-\vec{k}_2,-\vec{p})^* \big) \nonumber\\
&& + (e^ { -i \tau'  (k' + | \vec{p}-\vec{k}'| + p)} -  e^ { -i \tau'  (k' + | \vec{p}-\vec{k}'| - p)} )  \big ( \mathcal{G}_2 (\vec{k}_1,\vec{k}_2,\vec{k}') + \mathcal{G}_2 (\vec{k}_2,\vec{k}_1,\vec{k}') \big ) \nonumber\\
&&+ (e^ { -i \tau'  (k' - | \vec{p}-\vec{k}'| + p)} -  e^ { -i \tau'  (k' - | \vec{p}-\vec{k}'| - p)} ) \big ( \mathcal{G}_3(\vec{p}-\vec{k}',\vec{k}_1,\vec{k}_2) + \mathcal{G}_3 (\vec{p}-\vec{k}',\vec{k}_2,\vec{k}_1) \big) \nonumber\\
&&+ (e^ { -i \tau'  (-k' + | \vec{p}-\vec{k}'| +p)} -  e^ { -i \tau'  (-k' + | \vec{p}-\vec{k}'| - p)} ) \big ( \mathcal{G}_3(\vec{k}',\vec{k}_1,\vec{k}_2) + \mathcal{G}_3(\vec{k}',\vec{k}_2,\vec{k}_1) \big )  \nonumber\\
&&+ (e^ { -i \tau'  (- k' - | \vec{p}-\vec{k}'| - p)} -  e^ { -i \tau'  (-k' - | \vec{p}-\vec{k}'| + p)} )  \big( \mathcal{G}_2  (-\vec{k}_1,-\vec{k}_2,-\vec{k}')^* +\mathcal{G}_2   (-\vec{k}_2,-\vec{k}_1,-\vec{k}')^* \big ) \Bigg ] \nonumber\\
&&+ (2 \pi)^9 \delta(\vec{p}-\vec{k}_3)\big( \delta(\vec{k}' + \vec{k}_1)  \delta(\vec{k}_2+\vec{p} -\vec{k}') + \delta(\vec{k}_2 +\vec{k}') \delta(\vec{k}_1+\vec{p}-\vec{k}' ) \big ) \nonumber\\
&& \times \Bigg [  (e^ { -i \tau'  (-k' - | \vec{p}-\vec{k}'| + p)} -  e^ { -i \tau'  (k' + | \vec{p}-\vec{k}'| + p)} ) \mathcal{F}_1 (-\vec{p})+  (e^ { -i \tau'  (k' + | \vec{p}-\vec{k}'| - p)} -  e^ { -i \tau'  (-k' - | \vec{p}-\vec{k}'| - p)} ) \mathcal{F}_2 (-\vec{p}) \nonumber\\
&& +( e^ { -i \tau'  (k'+ | \vec{p}- \vec{k}'| +p)} - e^ { -i \tau'  (-k' - | \vec{p}- \vec{k}'| + p)}  )\mathcal{F}_2  (\vec{p})^* +  (e^ { -i \tau'  (-k' - | \vec{p}-\vec{k}'| - p)} -  e^ { -i \tau'  (k' + | \vec{p}-\vec{k}'| - p)} ) \mathcal{F}_1  (\vec{p})^* \Bigg ] \nonumber\\
&& -2 i (2 \pi)^9 \delta(\vec{p}-\vec{k}_3) \big( \delta(\vec{k}' + \vec{k}_1)  \delta(\vec{k}_2+\vec{p} -\vec{k}') + \delta(\vec{k}_2 +\vec{k}') \delta(\vec{k}_1+\vec{p}-\vec{k}' ) \big )\sin{\tau' (k'+|\vec{p}-\vec{k}| +p) }\nonumber\\
&&  + (\vec{p} \leftrightarrow -\vec{k}' ) + (\vec{p} \leftrightarrow \vec{k}' - \vec{p} ). \nonumber\\
&&+\text{cyclic perms}.
\end{eqnarray}
We now substitute the expression (\ref{B7}) into Eq.\ (\ref{secondterm}) and then
into Eq.\ (\ref{5.4}), and then simplify using the decompositions (\ref{decom}).
This yields the remaining pieces of our expression (\ref{finalresult}) for the
bispectrum: the remaining part of the vacuum term ${\cal B}_{0,{\rm
    dr}}$, the piece
${\hat {\cal A}}_{II}$ of the term ${\cal B}_{II}$, and the piece
${\hat {\cal A}}_{IV}$ of the term ${\cal
  B}_{IV}$. It also generates an additional term, from the time integral of the
third last line of Eq.\ (\ref{B7}), which cancels the
contribution (\ref{B.6}).


\providecommand{\href}[2]{#2}\begingroup\raggedright\endgroup

\end{document}